%% file: sensitivity.tex
\documentclass[10pt]{article}
\usepackage{amsmath}
\usepackage{amssymb}

\usepackage{graphicx}

\usepackage{cite}

\usepackage{color} 

\usepackage{setspace} 
\usepackage[labelfont=bf,labelsep=period,justification=raggedright]{caption}

\makeatletter
\renewcommand{\@biblabel}[1]{\quad#1.}
\makeatother

\date{}

\usepackage{rotating}

\newcommand{\ms}{\medskip}
\newcommand{\bs}{\bigskip}
\newcommand{\noi}{\noindent}

\newcommand{\dsty}{\displaystyle}
\newcommand{\txty}{\textstyle}

\newcommand{\be}{\beta}

\newcommand{\sig}{\sigma}
\newcommand{\Sig}{\Sigma}

\newcommand\la{\langle}
\newcommand\ra{\rangle}

\newcommand{\bl}[1]{\mathbf{#1}}

\newcommand{\bsu}{\mathop{\txty{\sum}}\limits}
\newcommand{\pro}{\mathop{\txty{\prod}}\limits}

\newcommand{\intl}{\int\limits}

\def\Xint#1{\mathchoice
    {\XXint\displaystyle\textstyle{#1}}%
    {\XXint\textstyle\scriptstyle{#1}}%
    {\XXint\scriptstyle\scriptscriptstyle{#1}}%
    {\XXint\scriptscriptstyle\scriptscriptstyle{#1}}%
    \!\int}
\def\XXint#1#2#3{\setbox0=\hbox{$#1{#2#3}{\int}$}
    \vcenter{\hbox{$#2#3$}}\kern-0.5\wd0}
\def\bint{\Xint-}
\def\dashint{\Xint{\raise4pt\hbox to7pt{\hrulefill}}}
\def\dashiint{\bint\kern-0.15cm\bint}
\newcommand{\ovl}[3]{\int_{#1}^{#2}\kern-#3pt\raise4pt\hbox to7pt{\hrulefill}\ }
\newcommand{\ovll}[3]{\intl_{#1}^{#2}\kern-#3pt\raise4pt\hbox to7pt{\hrulefill}\ }
\newcommand{\tvl}[2]{\iint_{#1}\kern-#2pt\raise4pt\hbox to7pt{\hrulefill}\ }

\newcommand{\bye}{\end{document}}
\newcommand{\Ca}{Ca$^{2+}$} 
\newcommand{\tCa}{\text{[Ca$^{2+}$]}}
\newcommand{\tpk}{\text{t}_{\text{i}}} 
\newcommand{\tipk}{\text{t}^{\text{peak}}_i} 
\newcommand{\tpkca}{\text{t}^{\text{peak}}_{\text{\Ca}}} 
\newcommand{\tpkrho}{\text{t}^{\text{peak}}_{\text{Rho}}} 
\newcommand{\ssc}{\text{s}} 
\newcommand{\gbj}{\text{g}_{\text{obj};i}} 
\newcommand{\Gbj}{\text{G}_{\text{obj};i}} 
\newcommand{\bbk}{\bar{\bl{k}}} 
\newcommand{\yimax}{y^{\text{max}}_i} 

\begin{document}

\begin{flushleft}
{\Large
\textbf{Identification of the key parameters in a mathematical   
model of PAR1-mediated  signaling in endothelial cells}
}
\\
Leonardo Lenoci$^{1}$, 
Heidi E. Hamm$^{1}$, 
Emmanuele DiBenedetto$^{2,\ast}$
\\
\bf{1} Department of Pharmacology, 
Vanderbilt University, Nashville, Medical Center, TN 37232, USA
\\
\bf{2} Department of Mathematics, 
Vanderbilt University, Nashville, TN 37240, USA
\\
$\ast$ E-mail: em.diben@vanderbilt.edu
\end{flushleft}

\section*{Abstract}
Biophysical models are often populated by 
a large number of input parameters that are difficult to predict or  measure experimentally.
The validity and robustness of a given model can 
be evaluated by a sensitivity test
to its input parameters. In this study, we performed  
local (based on a Taylor-like method) and global sensitivity (based on 
Monte Carlo filtering techniques) analyses 
of a previously derived  PAR1-mediated activation model of 
endothelial cells. This activation model previously demonstrated that     
peptide-activated PAR1 has a 
different receptor/G-protein binding affinity that 
favors $\mathrm{G\alpha_q}$ activation over $\mathrm{G\alpha_{12/13}}$ by 
approximately 800-fold. 
Interestingly, the present 
study  shows that the parameter  regulating the binding rate 
of activated PAR1 to  $\mathrm{G\alpha_{12/13}}$ is indeed 
important to obtain the expected $\mathrm{RhoGTP}$ response. Moreover,
we show that the parameters representing the rate of PAR1 deactivation and 
the rate of PAR1 binding to $\mathrm{G_q}$, are the 
most important parameters in the system.  Finally, we illustrate that 
the  kinetic model considered in this study is  robust and
we provide complementary insights into the biological 
meaning and importance  of its kinetic parameters.

\section*{Introduction}
Mathematical models of biophysical phenomena often involve 
a large number of input 
parameters~\cite{modules,lenoci_2010,diamond_2008,mclaughlin_2005,andreucci_2003},
such as reaction 
rates and initial concentrations, that  must be either measured experimentally or inferred 
from similar biological systems. Moreover, experimentally measured parameters 
carry uncertainties due to experimental 
limitations, statistical analysis and different experimental 
conditions.
One of the major goals in systems biology is to estimate how sensitive 
a computational model is to variations of its parameters 
and  study the effect of the parameter uncertainties 
on the model response.  Sensitivity analysis aims at 
determining which parameters have the most influence 
on a predicted system behavior\cite{sun_2005,wolkenhauer_2003,
deisboeck_2009,shen_2010}. When the notion of ``influence'' 
is made quantitatively precise, sensitivity analysis 
can constitute a reliable robustness test for computational 
models~\cite{Sensitivity}.

\begin{figure*}[!ht]
\begin{center}
\includegraphics[width=0.6\textwidth]{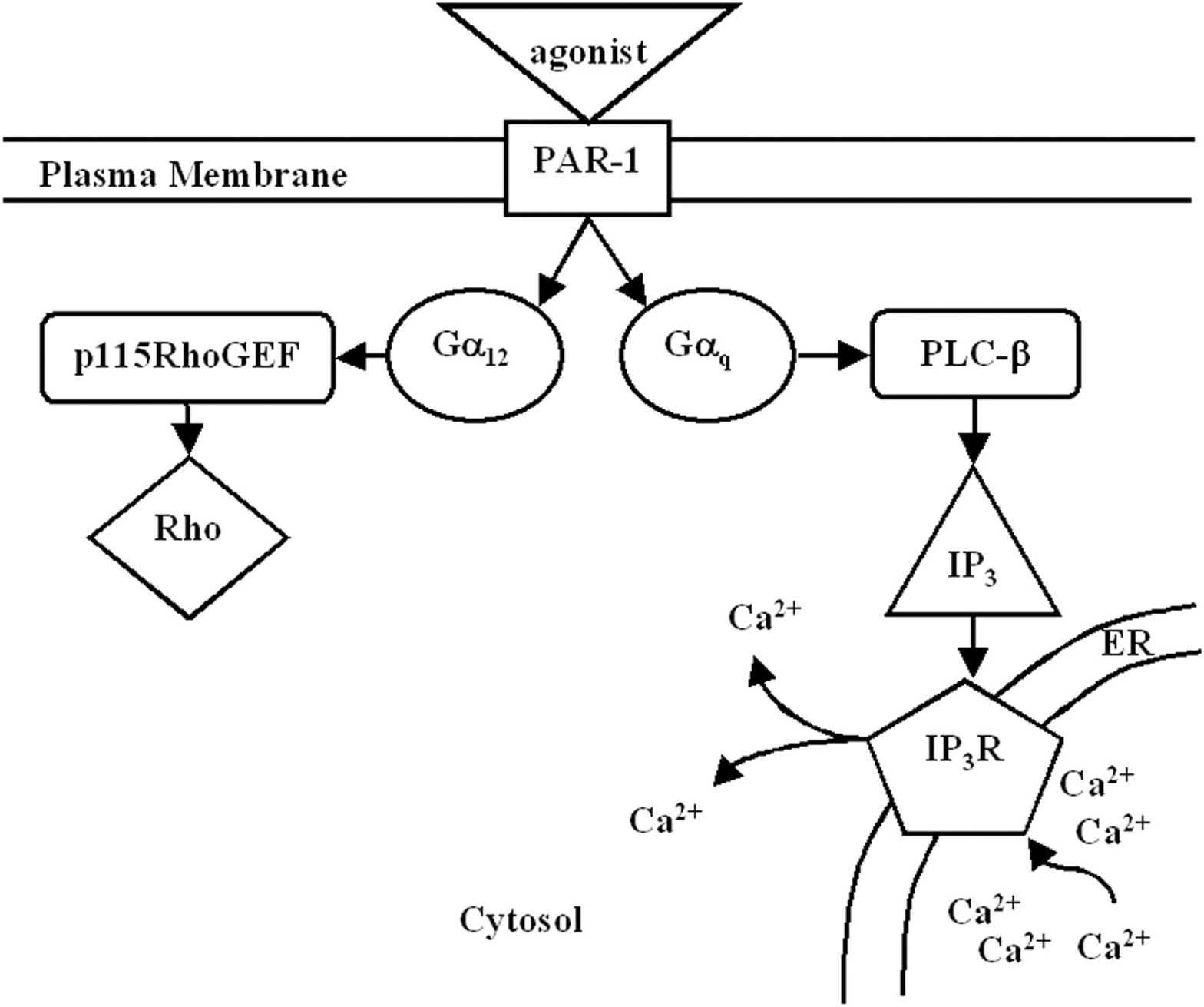}
\end{center}
\caption{{\bf Pictorial representation of  
PAR1-mediated signaling pathways in 
endothelial cells}. Upon cell stimulation by a PAR1 agonist peptide such as 
\mbox{SFLLRN}, PAR1 simultaneously activates 
the two classes of G proteins, $\mathrm{G_q}$ and  $\mathrm{G_{12/13}}$.
The $\alpha$ 
subunit of the ${\rm G_q}$ protein activates the $\beta$ isoforms of phospholipase C (${\rm PLC\beta}$),
which hydrolyze the lipid phosphatidylinositol 4,5-bisphosphate (PIP2) to generate the 
second messengers inositol 1,4,5-trisphosphate (IP3). IP3 then regulates the activity 
of the inositol trisphosphate receptors (${\rm IP_3R}$) on the 
surface of the endoplasmic reticulum (ER) allowing the rapid release of ${\rm Ca^{2+}}$ 
into the cytoplasm. Simultaneously, the $\alpha$
subunit of the ${\rm G_{12/13}}$ protein activates the small \mbox{GTPase} Rho
to promote cytoskeletal changes. Reproduced from~\cite{mclaughlin_2005}.}
\label{fig:path}
\end{figure*}

In this study, we performed sensitivity analysis on 
the mathematical model  of PAR1-mediated activation of 
endothelial cells of \cite{mclaughlin_2005}. 
A schematic representation of the main
signaling pathways analyzed in~\cite{mclaughlin_2005} is 
shown in Fig.~\ref{fig:path}.
A specific PAR1 agonist, \mbox{SFLLRN}, 
simultaneously activates two classes of G proteins: 
${\rm G_q}$  and   
${\rm G_{12/13}}$~\cite{coughlin_2000}. The $\alpha$ subunit 
of the ${\rm G_q}$ protein activates the $\beta$ 
isoforms of phospholipase C (${\rm PLC\beta}$), which 
hydrolyze the lipid phosphatidylinositol 4,5-bisphosphate 
(PIP2) to generate the second messengers inositol 
1,4,5-trisphosphate (IP3) and diacylglycerol 
(DAG)~\cite{rittenhouse_1983}. The second messenger IP3 regulates the 
activity of the inositol trisphosphate receptors 
(${\rm IP_3R}$) on the surface of the endoplasmic reticulum 
(ER), allowing the rapid release of ${\rm Ca^{2+}}$ 
into the cytoplasm~\cite{Platelets,keizer_1992}. 
Simultaneously, the $\alpha$ 
subunit of the ${\rm G_{12/13}}$ protein activates the 
small \mbox{GTPase} RhoA, known to promote cytoskeletal 
changes~\cite{burridge_1996}. 

The signaling pathways described above are also common to 
a number of transduction systems mediated by G-protein 
coupled receptors in other cell types. We have recently 
published a model of PAR1 signaling in platelets 
(\cite{Platelets}). 

In \cite{mclaughlin_2005}, several kinetic parameters  
had to be inferred because they were unknown and/or difficult 
to measure. In this work, we analyzed 
the influence of the chosen parameters to 
the system's output. We performed  a sensitivity test 
based on the Taylor expansion of the output functions around the 
chosen input parameters (local analysis) and a sensitivity test 
based on Monte Carlo sampling techniques of the 
system parameters (global analysis). Our results show that 
although the two techniques implemented are conceptually 
different, they lead, at least qualitatively, to consistent 
results. We show that the system is not sensitive to the 
majority of the parameters chosen in~\cite{mclaughlin_2005} 
and we identify three important nodes in the model.
These results represent a further test for the 
validity of the model in \cite{mclaughlin_2005}. Additionally, this analysis  will 
help design more refined computational models, and will 
have  the ultimate goal of identifying  influential  
parameters and key signaling nodes with possible 
applications in biology and medicine. 
\section*{Methods}
\subsection*{Sensitivity Analysis by Taylor Expansion}
Consider a mathematical  model defined by the following 
$q$-dimensional system of ordinary differential 
equations (ODEs) with initial conditions $\mathbf{y}_o$,
\begin{equation}
\left\{
\begin{array}{ll}
\dot{\bl{y}}&=f(\bl{y},\bl{k})\\
\bl{y}_o&=\bl{y}(0),
\end{array}\right.
\label{eq:sys}
\end{equation}
\input tableI.tex

where $\bl{y}=(y_1,\dots,y_q)$ is the $q$-dimensional 
vector of  system states and $\mathbf{k}=(k_1,\dots,k_n)$ 
is the  $n$-dimensional vector of input parameters. 
A simple estimate of the effect of the uncertainties 
in $\mathbf{k}$ on $\mathbf{y}$, that is, the variation 
of $\mathbf{y}$ caused by a variation $\Delta\mathbf{k}$ 
in $\mathbf{k}$ is
\begin{equation*}
\Delta \bl{y}(t,\bl{k})=\bl{y}(t,\bbk+\Delta\bl{k})
-\bl{y}(t,\bbk),
\end{equation*}
where $\mathbf{y}(t,\bbk)$ is a solution 
of~(\ref{eq:sys}) at time $t$ with parameters 
$\bl{k}=\bbk$. Assuming 
$|\Delta\bl{k}|\ll1$, by the Taylor's 
expansion in the variables $\mathbf{k}$ 
about $\bbk$, 
\begin{equation}\label{eq:taylor}
\Delta\bl{y}(t,\bl{k})= \bl{D}(t,\bbk)\Delta\bl{k} 
+\bl{O}(|\Delta\bl{k}|^2) 
\end{equation}
where $\bl{O}(|\Delta \bl{k}|^2)$ is a $q$-dimensional 
vector infinitesimal of order no less than $2$ with respect 
$|\Delta\bl{k}|$, and $\bl{D(t,\bbk)}$ is the $n\times q$ 
matrix of the partial derivatives at $\bbk$ of entries
\begin{equation*}
\bl{D}(t,\bbk)=\left(\frac{\partial y_i}{\partial k_j}
(t,\bl{k})\right)\Biggm|_{\bl{k}=\bbk}\qquad
\begin{array}{ll}
i=1,\dots,q\\
j=1,\dots,n
\end{array}
\end{equation*} 
Then one takes a dimensionless version of $\bl{D}(t,\bbk)$ as a 
measure of the sensitivity of the system at time $t$, with respect 
to the set of parameters $\bl{k}$. Precisely one introduces
a {\it sensitivity matrix} $\bl{S}(t,\bbk)$ of entries
\begin{equation}\label{eq:matrix_s}
S_{ij}(t,\bbk)=\frac{\partial\ln(y_i)}{\partial\ln(k_j)}
\Bigg|_{\bl{k}=\bbk}
=\frac{\bar{k}_j}{y_i(t,\bbk)}
\frac{\partial y_i(t,\bl{k})}{\partial k_j}\Bigg|_{\bl{k}=\bbk}
\end{equation} 
and takes $S_{ij}(t,\bbk)$ as a measure of the 
sensitivity of the 
system state $y_i(t,\bl{k})$ at time $t$ with respect to the kinetic 
parameter $k_j$ about the nominal values $\bbk$. In the 
context of the kinetic model  in \cite{mclaughlin_2005} 
we selected the states
\begin{equation}\label{eq:states}
\begin{aligned}
y_i(t,\bl{k})&=\tCa(t,\bl{k})\\
y_i(t,\bl{k})&=\text{[RhoGTP]}(t,\bl{k})
\end{aligned}
\end{equation}
and computed $S_{ij}(\tpk,\bbk)$ at time 
$\tpk$ where the nominal $\tCa(t,\bbk)$ and 
$\text{[RhoGTP]}(t,\bbk)$ attain their maximum 
values, e.g.,
\begin{equation}\label{eq:time-pk}
y_i(\tpk,\bbk)=\max_t y_i(t,\bbk).
\end{equation}
The sensitivity coefficients $S_{ij}(\tpk,\bbk)$ for {\Ca} 
are reported in Table~\ref{tab:tableI} in columns two and three along those 
for {RhoGTP} in columns four and five. The partial 
derivatives $\partial y_i/\partial k_j$ 
were approximated by discrete differences with values of 
$\pm 5\%$ of the nominal values and indicated respectively 
with the symbols $S_{+}$  and $S_{-}$. 

\input tableII.tex

The total output of $\bl{y}(\cdot,\bl{k})$ over the 
average time course $T$ of the experiment, is
\begin{equation}\label{eq:int_states}
\bl{z}(\bl{k})=\int_0^T \bl{y}(t,\bl{k})dt.
\end{equation}
For the states in (\ref{eq:states}) we chose in our simulations 
$T=600\ssc$. The vector $\bl{z}(\bl{k})$ is 
independent of time and 
it can be expanded in Taylor's series 
with respect to $\bl{k}$, as in (\ref{eq:taylor}) with 
$\bl{D}(t,\bbk)$ replaced by its time integral 
over $(0,T)$. The sensitivity coefficients of $\bl{z}(\bl{k})$ 
about the nominal values $\bbk$ are
\begin{equation}\label{eq:matrix_sig}
\Sig_{ij}(\bbk)=\frac{\bar{k}_j}{z_i(\bbk)}
\frac{\partial z_i}{\partial k_j}
(\bl{k})\Biggm|_{\bl{k}=\bar{\bl{k}}}.
\end{equation}
The sensitivity coefficients for {\Ca} are reported 
in Table~\ref{tab:tableII} in columns two and three  along those 
for {RhoGTP} in columns four and five. The partial 
derivatives $\partial y_i/\partial k_j$ 
were approximated by discrete differences with values of 
$\pm 5\%$ of the nominal values and indicated respectively 
with the symbols $\Sig_{+}$  and $\Sig_{-}$.

The nominal values of the parameters $k_j$ used in the 
sensitivity analysis described above are given in Table~\ref{tab:tableVIII}. 
\input tableVIII.tex
\subsection*{Monte Carlo Filtering Sensitivity Analysis}
The method is based on estimating the uncertainties 
distributions $\bl{p}(\bl{k})=\{p_1(k_1),\dots,p_n(k_n)\}$ of the 
parameters $\bl{k}=(k_1,\dots,k_n)$, each ranging 
over the intervals
\begin{equation}\label{eq:ranges}
k_j\in\big(k_j^{\text{min}}\,,\, k_j^{\text{max}}\big), \quad 
j=1,\dots,n
\end{equation}
and cumulatively generating the probability measure 
\begin{equation*}
\bl{p(k)} d\bl{k}= p_1(k_1)\cdots p_n(k_n)\, dk_1\cdots dk_n
\end{equation*}
over the space of parameters 
\begin{equation*}
\pro_{j=1}^n\big(k_j^{\text{min}}\,,\, k_j^{\text{max}}\big).
\end{equation*} 
If these ranges and uncertainties distributions 
were known, regarding each of the $y_i(t,\bl{k})$ as 
a random variable depending on the random $\bl{k}$ 
one computes the mean 
\begin{equation*}
\la y_i(t)\ra=\int\cdots\int_{k_j^{\text{min}}}^{k_j^{\text{max}}}
y_i(t,\bl{k})\, \bl{p(k)}\,d\bl{k}
\end{equation*}
and, the variance 
\begin{equation*}
\sig^2_i(t)=\la y_i^2(t)\ra- \la y_i(t) \ra^2,
\end{equation*}
 where 
\begin{equation*}
\la y_i^2(t)\ra=\int\ldots\int_{k_j^{\text{min}}}^{k_j^{\text{max}}}
y_i^2(t,\bl{k})\,\bl{p(k)}\,d\bl{k}.
\end{equation*}
The variance is a measure of the influence, 
and relative importance, of the input $\bl{k}$ to 
the output $\bl{y}(t,\bl{k})$ (\cite{Sensitivity}).

In practice the process is implemented in a less 
quantitative way, by trading the information coming 
from the variances $\sig_i(t)$ with those coming from 
a biologically motivated {\it objective functional}, 
$g_{\text{obj}}$ depending on one or several $y_i(t,\bl{k})$~\cite{hornberger_1981}.  
Having chosen an objective functional $g_{\text{obj}}$, 
one identifies a biologically {\it acceptable range} 
for the objective function $g_{\text{obj}}$. 
For example  for a {\it threshold value} $g^{\text{thres}}$ 
of the objective functional $g_{\text{obj}}$ one might define~\cite{hornberger_1981}
\begin{equation}\label{eq:obj_range}
\begin{array}{ll}
g_{\text{obj}}\le g^{\text{thres}} \quad&
\text{ as the acceptable range}\\
g_{\text{obj}}>g^{\text{thres}} \quad&
\text{ as the non acceptable range}.
\end{array}
\end{equation}
Then for each $k_j$ one determines the probability 
distributions $f_1(k_j)$ and $f_2(k_j)$  
of those values of $k_j$ that output the system in 
the acceptable or non acceptable range respectively. For 
each of these, and for each $k_j$, calculate the 
cumulative frequency distributions
\begin{equation}\label{eq:cumul}
cf_{\ell}(k_j)=\frac{\dsty 
\int_{k_j^{\text{min}}}^{k_j}
f_{\ell}(\eta)d\eta}{\dsty\max_{[k_j^{\text{min}}, k_j^{\text{max}}]}
f_{\ell}(k_j)}, \qquad\ell=1,2
\end{equation}
and calculate the Kolmogorov-Smirnov coefficients $d_{1,2}(k_j)$ 
by the maximum distance function
\begin{equation}\label{eq:kolm-smir}
d_{1,2}(k_j)=\sup_{k_j\in
[k_j^{\text{min}}, k_j^{\text{max}}]}|cf_1(k_j)-cf_2(k_j)|.
\end{equation}
These coefficients are taken as a measure of the relative 
importance of each parameter $k_j$ on the model output. 
The larger the value of $d_{1,2}(k_j)$, the more 
important is $k_j$ in producing the pre-defined system output~\cite{hornberger_1981,wolkenhauer_2003,sun_2005,deisboeck_2009}. 

In the context of the kinetic model of \cite{mclaughlin_2005} the 
method is implemented as follows:

\bs

\noi{\bf 1.\ } Select nominal values 
$\bbk=\big(\bar{k}_1,\dots,\bar{k}_n\big)$ 
as those originally adopted in the model of 
\cite{mclaughlin_2005} (see Table~\ref{tab:tableVIII}). As ranges in (\ref{eq:ranges}) 
we take intervals spanning from $1/10$ to $10$ 
times  these nominal values, e.g., 
\begin{equation*}
k_j^{\text{max}} =10 \bar{k}_j,\quad\text{ and }\quad 
k_j^{\text{min}}={\txty\frac1{10}}\bar{k}_j.
\end{equation*}
On a $\log_{10}$ scale these are symmetric intervals about 
$\log_{10}\bar{k}_j$.
\bs

\noi{\bf2.\ } Generate $M$ random $n$-tuples 
of numbers 
\begin{equation*}
\bl{k}_m=(k_{1,m},\dots,k_{n,m})\quad\text{ for }\> 
 m=1,\dots,M
\end{equation*}
by uniformly sampling the $\ln k_j$ from their respective 
symmetric uncertainty ranges defined above. In the simulations 
we used $M=10,000$. Different sampling distributions 
(e.g. gaussian, exponential) were seen not to qualitatively affect the results.
\input tableIII.tex

\bs

\noi{\bf3.\ }  Solve the system (\ref{eq:sys}) for each random 
choice of $\bl{k}_m$, and compute the functions 
$\bl{y}(t,\bl{k}_m)$. Solve also 
(\ref{eq:sys})  for the nominal values $\bl{k}=\bbk$ 
to get the functions $\bl{y}(t,\bbk)$. Then  
for each $i=1,\dots,q$ introduce two kinds of objective 
output functions of the $\text{m}th$ trial as follows:
\input tableIV.tex

\ms

\noi{\it Objective Function at Times $\tpk$:}  
\begin{equation}\label{eq:gbj}
\gbj(\tpk;m)=[y_i(\tpk,\bl{k}_m)-
y_i(\tpk,\bbk)]^2
\end{equation}
where $m\in\{1,\dots,M\}$ is the m$th$ random trial described 
in the previous step and the times $\tpk$ are defined in 
(\ref{eq:time-pk}). For the states in (\ref{eq:states}), 
this function measures the variations of 
the of $\tCa(t_{\text{\Ca}},\bl{k})$ and 
$\text{[RhoGTP]}(t_{\text{Rho}},\bl{k})$, 
from their nominal maximum values. 
(see Table~\ref{tab:tableIII} columns two and three respectively).    

\ms

\noi{\it Objective Function for Total Time:}  

\ms

Let $\bl{z}(\bl{k})$ be the total state output 
as in (\ref{eq:int_states}) and for $i=1,\dots,q$ set
\begin{equation}\label{eq:Gbj}
\Gbj(m)=[z_i(\bl{k}_m)-z_i(\bbk)]^2 
\end{equation}
where $m\in\{1,\dots,M\}$ is the m$th$ random trial. 
For the states in (\ref{eq:states}), the function 
$\Gbj(m)$ measures the perturbation of the total outputs of 
{\Ca} and \mbox{RhoGTP} over the 
time course $T$ of the experiment, from 
their nominal values. 
The results are compared in Table~\ref{tab:tableIV} in columns two and three.    

\bs

\noi{\bf4.\ } Introduce threshold values
\begin{equation}\label{eq:thresh}
\begin{aligned}
\gbj^{\text{thres}}(\tpk)&=\frac1M\bsu_{m=1}^M \gbj(\tpk;m)\\
\Gbj^{\text{thres}}&=\frac1M\bsu_{m=1}^M \Gbj(m).
\end{aligned}
\end{equation}
The m$th$ random trial and its parameters 
$\bl{k}_m$ are deemed {\it acceptable} according 
to the criterion in (\ref{eq:obj_range}) for each 
of these objective functions and their respective 
threshold values. On the basis of this classification, 
generate the probability distribution functions 
$f_1(k_j)$ and $f_2(k_j)$ of acceptable and unacceptable 
values, relative to $\gbj(\tpk;m)$ and $\Gbj(m)$ respectively. 

\bs

\noi{\bf5.\ } 
Calculate the cumulative frequency distributions 
$cf_{\ell}(k_j)$ as in (\ref{eq:cumul}) relative to 
each objective function, and the corresponding 
Kolmogorov-Smirnov coefficient $d_{1,2}(k_j)$ 
as in (\ref{eq:kolm-smir}). 
The larger the value of $d_{1,2}(k_j)$ the higher 
the sensitivity of the system to the variation of the 
corresponding   parameter. 

\bs

All the calculations were performed on a 
\mbox{MATLAB}~(R2009b, \mbox{The Mathworks},
Natick, MA) platform.
\subsection*{Times to Peak and Peak Values}\label{S:peak}
\input tableV.tex

\input tableVI.tex

According to (\ref{eq:time-pk}) the times $t_i$ are those 
at when the nominal states $y_i(\cdot, \bbk)$ achieve 
their peak value. The states $y_i(\cdot,\bl{k})$ however 
achieve their peak values at times $\tipk$ which, in general 
differ from $t_i$ and are in general functions 
of $\bl{k}$.  Set
\begin{equation}
\yimax(\bl{k})=y_i\big(\tipk(\bl{k}),\bl{k}\big)=\max_t y_i(t,\bl{k}).
\label{eq:tpk-peak}
\end{equation}
Dose responses for different concentrations 
of agonists are measured at $\tipk$. In practice, for 
a set of parameters $\bl{k}$, which in general is 
unknown, this is done by recording 
the experimental outputs $\yimax(\bl{k})$. For small variations 
of $\bl{k}$ about its nominal vector $\bbk$, the sensitivity 
of $\yimax(\bl{k})$ could be theoretically ``measured'' 
as in (\ref{eq:matrix_s}), by the sensitivity matrix of entries
\begin{equation*}
\begin{aligned}
S_{ij}^{\text{peak}}(\bbk)=
\frac{\partial \ln \yimax}{\partial\ln(k_j)}\biggm|_{\bl{k}=\bbk}
=\frac{\bar{k}_j}{\yimax(\bbk)}\bigg\{
\frac{\partial y_i\big(\tipk(\bl{k}),\bl{k}\big)}{\partial k_j}
\quad +\frac{d y_i\big(\tipk(\bl{k}),\bl{k}\big)}{\partial dt} 
\frac{\partial \tipk(\bl{k})}{\partial k_j}
\bigg\}\bigg|_{\bl{k}=\bbk}
\end{aligned}
\end{equation*} 
This formula however requires the form of the functions 
$\bl{k}\to \tipk(\bl{k})$, which are in general 
not know. Alternatively, the analysis can be carried 
by the filtering method, by introducing two new objective 
functions:
\begin{equation}\label{eq:obj-pk}
\begin{aligned}
F^{\text{peak}}_{Y_i}(m)&=\big[\yimax(\bl{k}_m)- y_i(t_i,\bbk)\big]^2\\
T^{\text{peak}}_{Y_i}(m)&=\big[\tipk(\bl{k}_m)-t_i\big]^2
\end{aligned}
\end{equation}
where the vector $\bl{k}_m$ is the output of the 
m$th$ random trial. Threshold values and distribution 
functions can be defined and determined as above.  
For $y_i=\tCa$ and $y_i=\text{[RhoGTP]}$, denote their peak 
times by $\tpkca$ and $\tpkrho$ respectively, and set 
\begin{equation}\label{eq:value-pk}
\begin{aligned}
\tCa^{\text{max}}(\bl{k})&=\tCa\big(\tpkca(\bl{k}),\bl{k}\big)\\
\text{[RhoTGP]}^{\text{max}}(\bl{k})&=
\text{[RhoGTP]}\big(\tpkrho(\bl{k}),\bl{k}\big).
\end{aligned}
\end{equation}
Then the first of (\ref{eq:obj-pk}) measures the variations 
of the peak values of $\tCa$ and \mbox{[RhoGRP]} from their 
peak nominal values, whereas the second of (\ref{eq:obj-pk}) 
measures the variation of their times to peak 
$\tpkca$ and $\tpkrho$ respectively, from their 
nominal values $t_{\text{\Ca}}$ and $t_{\text{RhoGTP}}$. 
The results are reported in Tables~\ref{tab:tableV} 
and \ref{tab:tableVI} respectively.    
\section*{Results and Discussion}\label{S:discussion}
We performed sensitivity analysis of a previously derived 
PAR1-mediated activation model of endothelial cells.
We used two different techniques, one based on the Taylor expansion of the 
system's output around the nominal values of its input parameters $k_j$ and the second 
based on Monte Carlo sampling techniques of the model parameters. 
The analysis based on the Taylor 
expansions (\ref{eq:taylor}) and leading to the 
sensitivity matrices $S_{ij}(t_i,\bbk)$ in (\ref{eq:matrix_s}) 
and $\Sig_{ij}(\bbk)$ in (\ref{eq:matrix_sig}), 
imposes two restrictions. 
The first is that $|\Delta\bl{k}|\ll1$ with the notion 
of ``smallness'' depending on a predefined notion 
of smallness of $|\Delta\bl{y}|$. The second is that 
these matrices measure the relative variation of $y_i$ 
and $z_i$ with respect to $k_j$, by keeping all the 
remaining parameters fixed, thereby neglecting the 
cumulative effects of $\Delta\bl{k}$. 
\begin{figure*}[!ht]
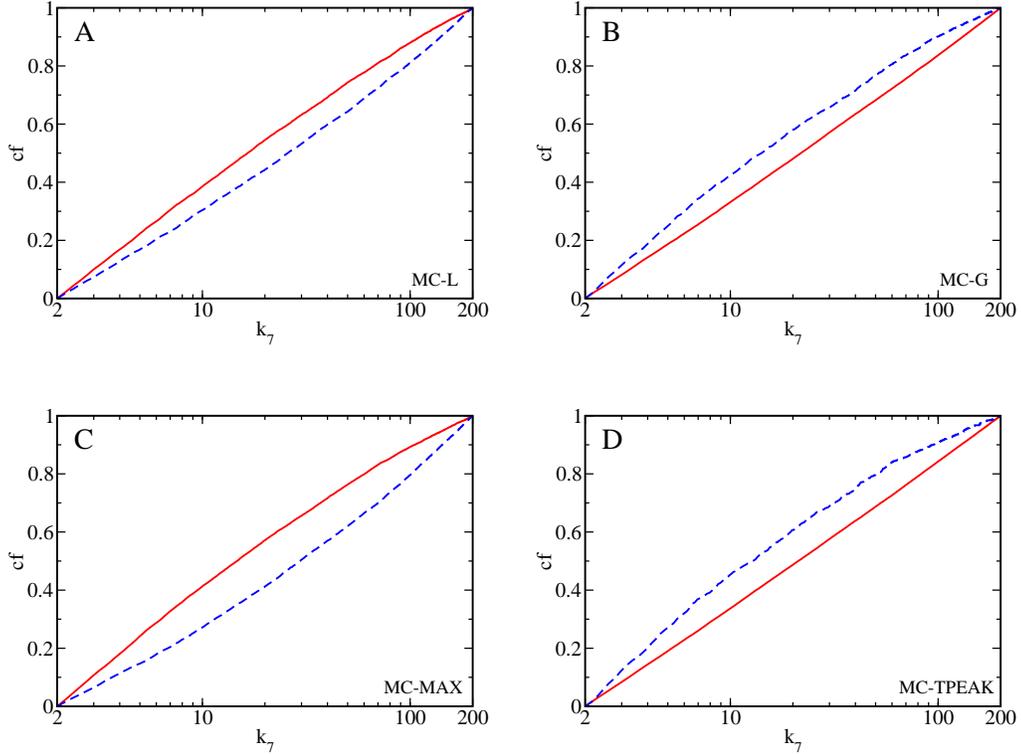

\begin{center}
\includegraphics[width=0.4\textwidth]{ca_k7_MC_L_log.eps}\hspace{0.5cm}
\includegraphics[width=0.4\textwidth]{ca_k7_MC_G_log.eps}\\
\vspace{0.8cm}
\includegraphics[width=0.4\textwidth]{ca_k7_MC_MAX_log.eps}\hspace{0.5cm}
\includegraphics[width=0.4\textwidth]{ca_k7_MC_TPEAK_log.eps}
\end{center}
\caption{
{\bf Cumulative distribution functions of 
the most influential kinetic parameter $k_7$ for the output $y_i=\mathrm{Ca^{2+}}$.}  
The red curves indicate the distributions relative to the  ``acceptable'' 
sets of parameters, whereas  the blue ones denote those corresponding to the 
``unacceptable'' set of parameters (see {\bf Methods}).
Panel~A: Monte Carlo filtering method results at the nominal peak time
$\mathrm{t_i}$. Panel~B: Monte Carlo filtering results for the entire time-course of the simulation,
$T=600~\mathrm{s}$. Panel~C: Monte Carlo filtering results  for the peak values $\mathrm{[Ca^{2+}]^{max}}$
of the output function. Panel~D: Monte Carlo filtering results for the peak times $\mathrm{t_{i}^{peak}}$ of the 
output function. The maximum distances between the acceptable and unacceptable distributions, 
$d_{1,2}$ are reported in Tables~\ref{tab:tableIII}--\ref{tab:tableVI} respectively and are 
defined in (\ref{eq:kolm-smir}).
The nominal peak time $\mathrm{t_i}$ is defined in (\ref{eq:time-pk})
and the objective functions used are respectively (\ref{eq:gbj}), (\ref{eq:Gbj}),
and (\ref{eq:obj-pk}) defined in \bf{Methods}.}
\label{fig:k7_ca}
\end{figure*}

The analysis based on the Kolmogorov-Smirnov test, does 
not require the range of $\bl{k}$ to be small, however it 
does require that the uncertainty distributions 
$\bl{p}(\bl{k})$ be known. The method leaves open 
the choice of the objective function. This on 
the one hand affords the flexibility in tailoring the 
objective function to specific experimental processes, 
and on the other hand allows non quantitative 
elements in the analysis. 
\begin{figure*}[!ht]
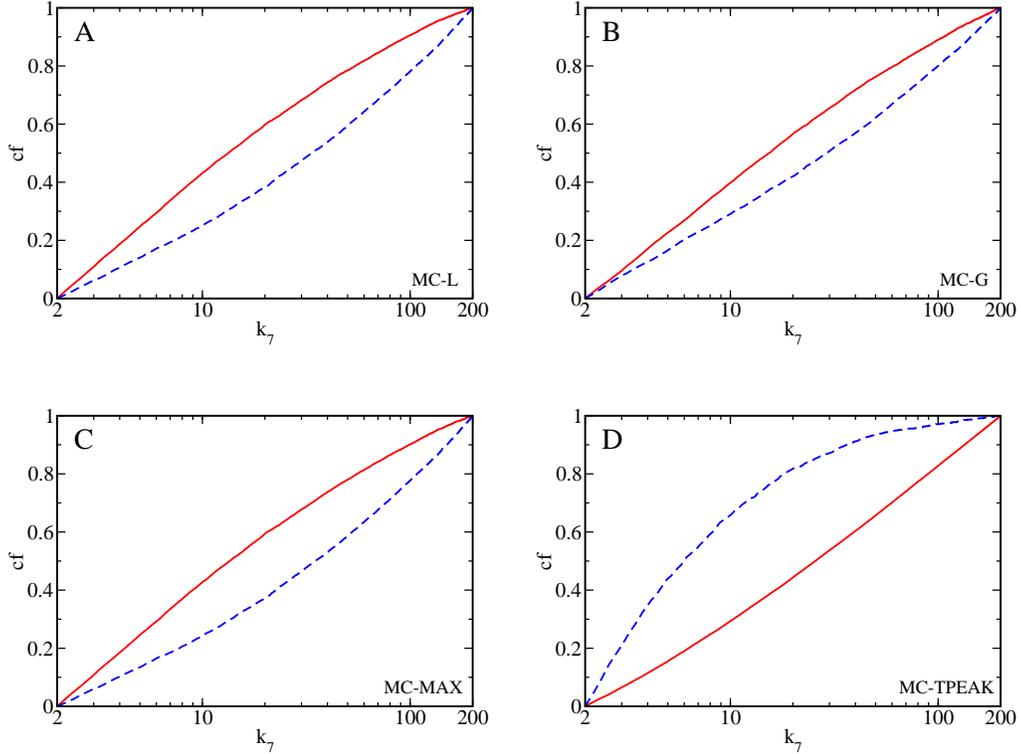

\begin{center}
\includegraphics[width=0.4\textwidth]{rho_k7_MC_L_log.eps}\hspace{0.5cm}
\includegraphics[width=0.4\textwidth]{rho_k7_MC_G_log.eps}\\
\vspace{0.8cm}
\includegraphics[width=0.4\textwidth]{rho_k7_MC_MAX_log.eps}\hspace{0.5cm}
\includegraphics[width=0.4\textwidth]{rho_k7_MC_TPEAK_log.eps}
\end{center}
\caption{
{\bf Cumulative distribution functions of 
the most influential kinetic parameter $k_7$ for the output $y_i=\mathrm{RhoGTP}$.}  
The red curves indicate the distributions relative to the  ``acceptable'' 
sets of parameters, whereas  the blue ones denote those corresponding to the 
``unacceptable'' set of parameters (see {\bf Methods}).
Panel~A: Monte Carlo filtering method results at the nominal peak time
$\mathrm{t_i}$. Panel~B: Monte Carlo filtering results for the entire time-course of the simulation,
$T=600~\mathrm{s}$. Panel~C: Monte Carlo filtering results  for the peak values $\mathrm{[RhoGTP]^{max}}$
of the output function. Panel~D: Monte Carlo filtering results for the peak times $\mathrm{t_{i}^{peak}}$ of the 
output function. The maximum distances between the acceptable and unacceptable distributions, 
$d_{1,2}$ are reported in Tables~\ref{tab:tableIII}--\ref{tab:tableVI} respectively and are 
defined in (\ref{eq:kolm-smir}).
The nominal peak time $\mathrm{t_i}$ is defined in (\ref{eq:time-pk})
and the objective functions used are respectively (\ref{eq:gbj}), (\ref{eq:Gbj}),
and (\ref{eq:obj-pk}) defined in \bf{Methods}.}
\label{fig:k7_rho}
\end{figure*}

The two methods being complementary, we performed 
a sensitivity analysis by using both of them, 
on the states in (\ref{eq:states})  arising from 
the mathematical model of \cite{mclaughlin_2005}. 

First we analyzed the sensitivity of {\Ca} and 
\mbox{RhoGTP} at their nominal peak values (\ref{eq:time-pk}) to 
variations of the parameters $\bl{k}$. Using the
Taylor expansion method, we computed the sensitivity 
coefficients $S_{ij}(\tpk,\bbk)$ introduced in 
(\ref{eq:matrix_s}) and reported them in Table~\ref{tab:tableI}. 
Using the Monte Carlo filtering method, starting 
from the objective function $\gbj(\tpk;m)$ 
introduced in (\ref{eq:gbj}), we computed the 
Kolmogorov-Smirnov coefficients by the distance function 
in (\ref{eq:kolm-smir}), and reported them in 
Table~\ref{tab:tableIII}.    
\begin{figure*}[!ht]
\begin{center}
\includegraphics[width=0.4\textwidth]{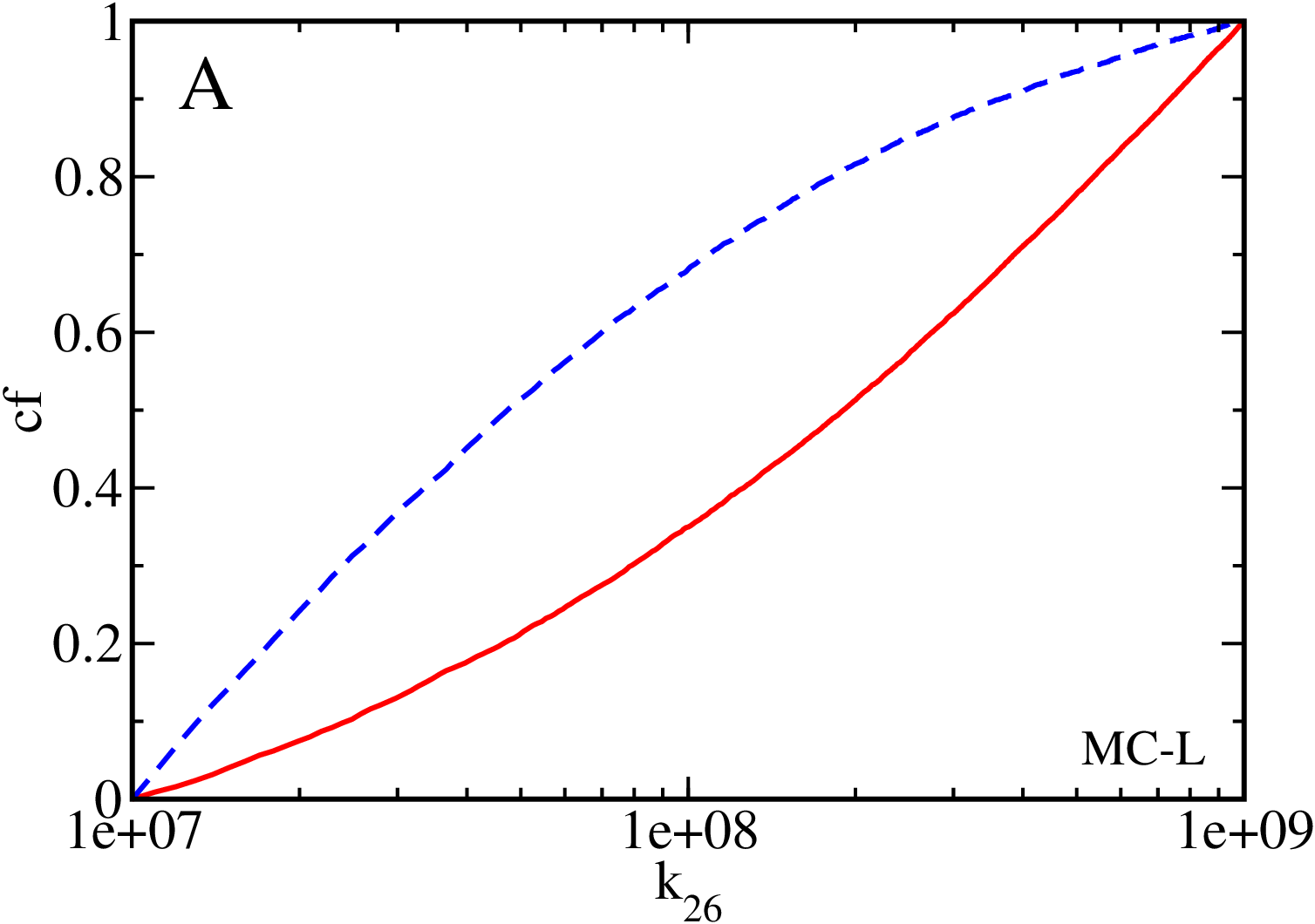}\hspace{0.5cm}
\includegraphics[width=0.4\textwidth]{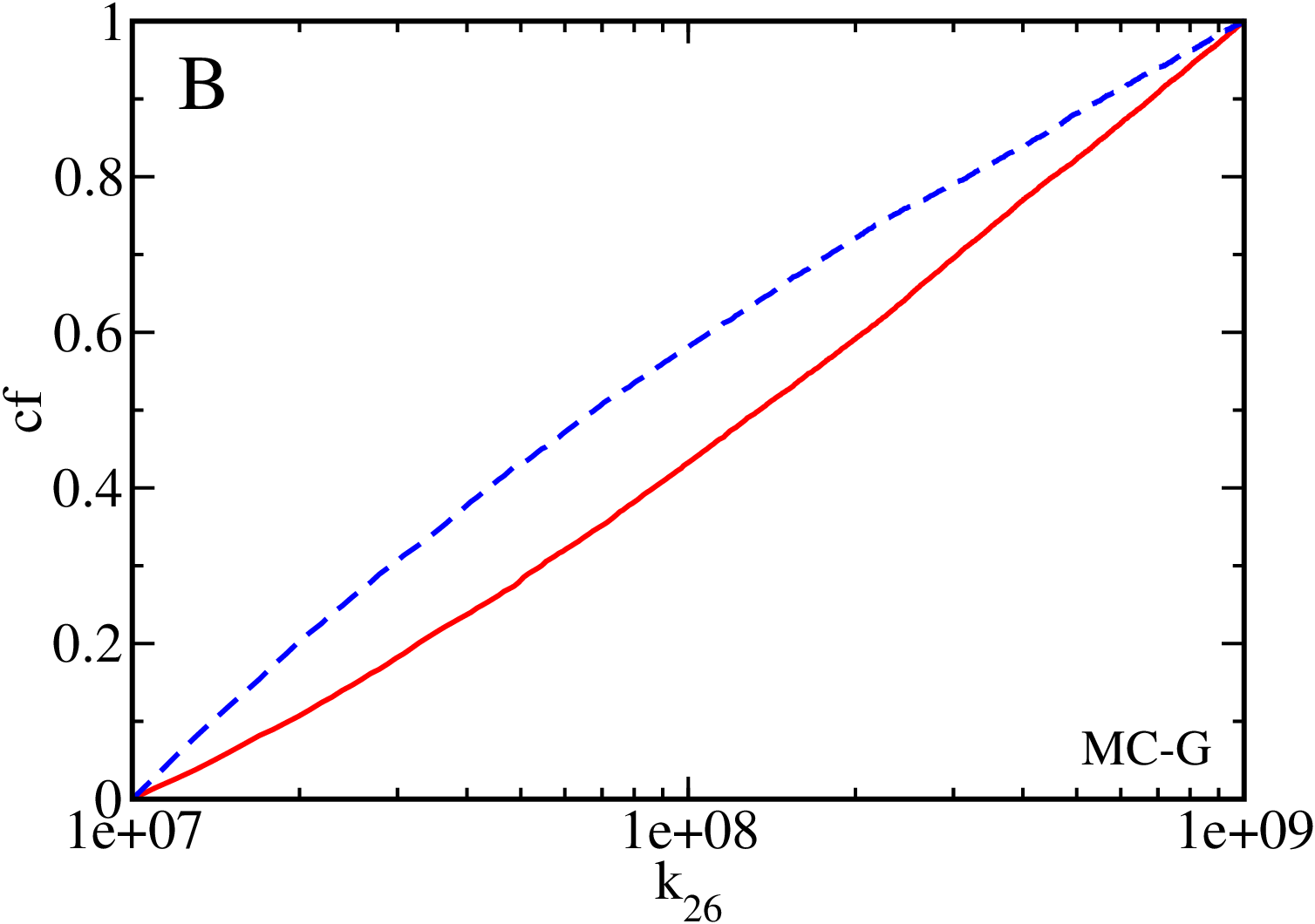}\\
\vspace{0.8cm}
\includegraphics[width=0.4\textwidth]{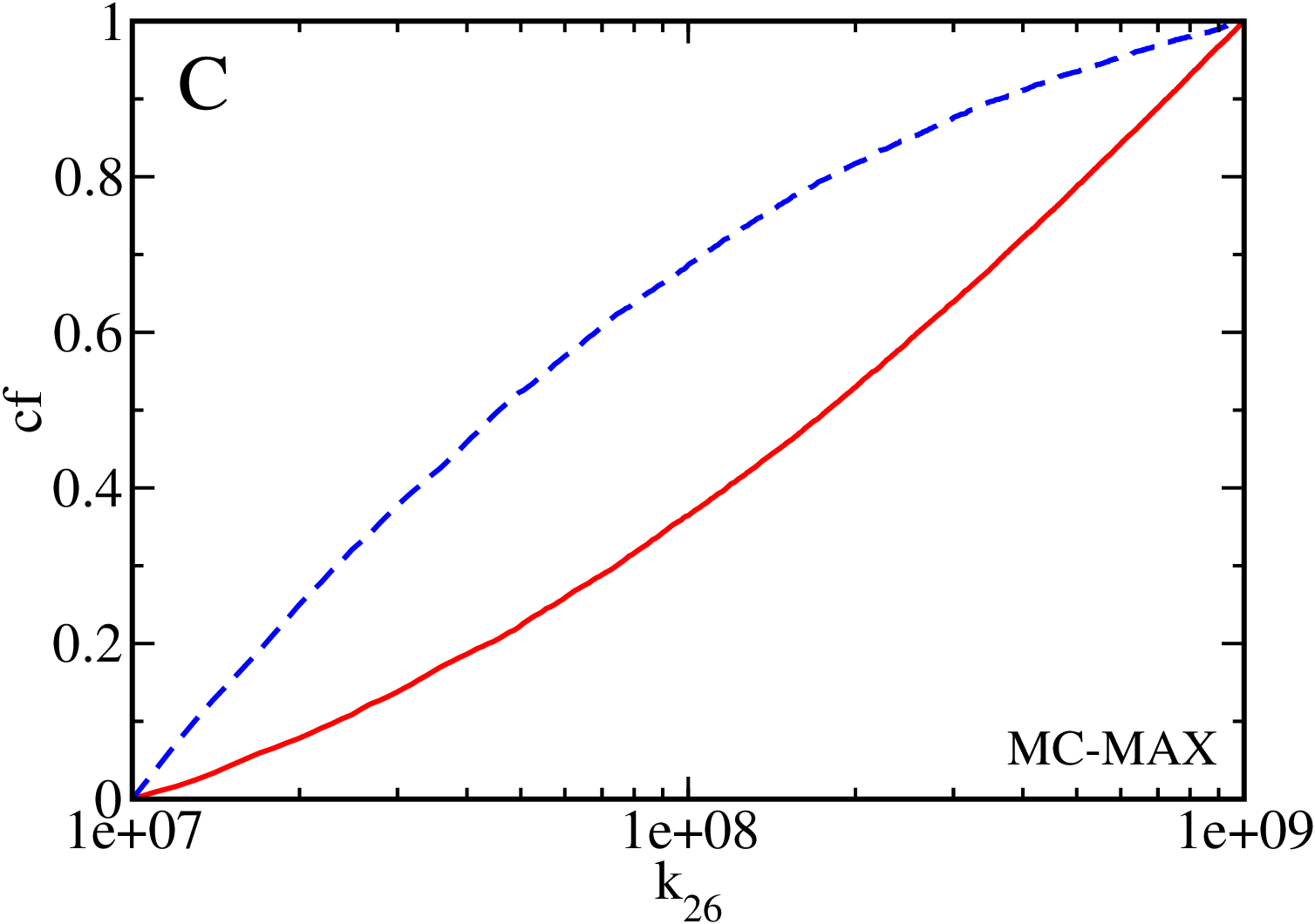}\hspace{0.5cm}
\includegraphics[width=0.4\textwidth]{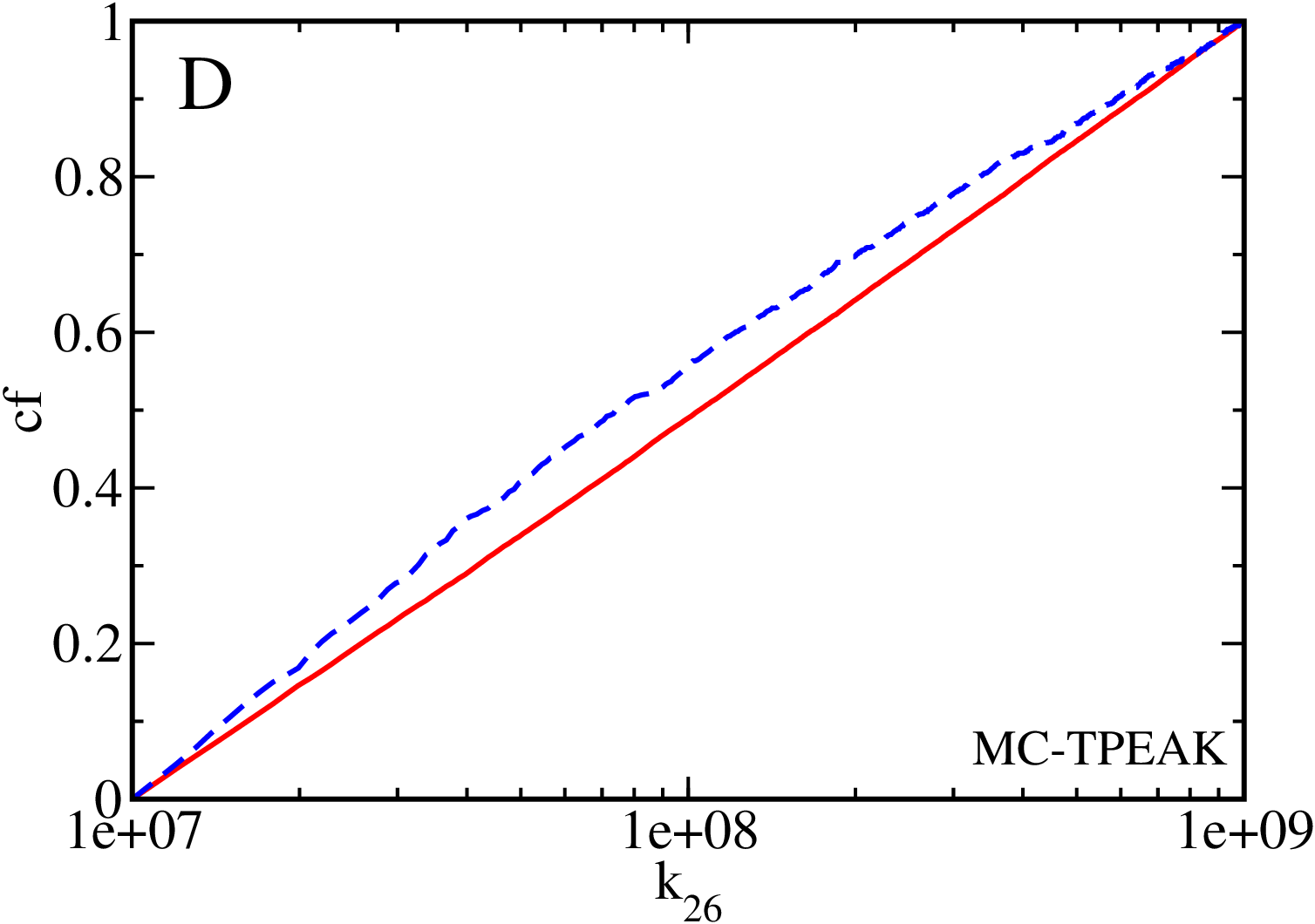}
\end{center}
\caption{
{\bf Cumulative distribution functions of 
the  influential kinetic parameter $k_{26}$ for the output $y_i=\mathrm{RhoGTP}$.}  
The red curves indicate the distributions relative to the  ``acceptable'' 
sets of parameters, whereas  the blue ones denote those corresponding to the 
``unacceptable'' set of parameters (see {\bf Methods}).
Panel~A: Monte Carlo filtering method results at the nominal peak time
$\mathrm{t_i}$. Panel~B: Monte Carlo filtering results for the entire time-course of the simulation,
$T=600~\mathrm{s}$. Panel~C: Monte Carlo filtering results  for the peak values $\mathrm{[RhoGTP]^{max}}$
of the output function. Panel~D: Monte Carlo filtering results for the peak times $\mathrm{t_{i}^{peak}}$ of the 
output function. The maximum distances between the acceptable and unacceptable distributions, 
$d_{1,2}$ are reported in Tables~\ref{tab:tableIII}--\ref{tab:tableVI} respectively and are 
defined in (\ref{eq:kolm-smir}).
The nominal peak time $\mathrm{t_i}$ is defined in (\ref{eq:time-pk})
and the objective functions used are respectively (\ref{eq:gbj}), (\ref{eq:Gbj}),
and (\ref{eq:obj-pk}) defined in \bf{Methods}.}
\label{fig:k26_rho}
\end{figure*}

Then we analyzed how parameters fluctuations affect  
the total response of {\Ca} and \mbox{RhoGTP} over 
the whole time course $T$ of the experiments, 
quantified by the integrated states (\ref{eq:int_states}). 
We computed first the sensitivity coefficients $\Sig_{ij}(\bbk)$ 
in (\ref{eq:matrix_sig}), and reported in Table~\ref{tab:tableII}. 
Then, starting from the objective function $\Gbj(m)$ 
introduced in (\ref{eq:Gbj}), we computed the Kolmogorov-Smirnov 
coefficients as in (\ref{eq:kolm-smir}) and 
reported in Table~\ref{tab:tableIV}.

Finally, by the same Monte Carlo filtering procedure 
 we investigated  the sensitivity of the peak values of 
{\Ca} and \mbox{RhoGTP} and their relative times to peak 
for cumulative variations of all the parameters 
$k_j$ in intervals  of 2 orders of magnitude with 
respect to $\bar{k}_j$. The results are in Table~\ref{tab:tableV} 
and Table~\ref{tab:tableVI} respectively. 

As a way of analyzing and comparing the results in these tables, we 
deemed  a parameter  important for a given state $y_i(t,\bl{k})$ 
if its fluctuations about the nominal values $\bbk$ produced 
sensitivity coefficients or Kolmogorov-Smirnov coefficients 
larger than $10\%$. Equivalently the state $y_i$ was deemed 
not to be sensitive to variations of those parameters $k_j$ for which 
the Taylor sensitivity coefficients or the Kolmogorov-Smirnov 
sensitivity coefficients were smaller than $10\%$. 

The two methods are conceptually different 
and hence the sensitivity coefficients $S_{ij}(t_i,\bbk)$ and 
$\Sig_{ij}(\bbk)$ introduced in   (\ref{eq:matrix_s}) and 
(\ref{eq:matrix_sig}) respectively, are not expected 
to be numerically similar to the Kolmogorov-Smirnov 
coefficients relative to the same processes. Nevertheless 
they exhibit similar qualitative results, in the sense that 
most of the parameters that are important by the Taylor 
expansion method, are likewise important by the Monte Carlo 
filtering method. We first discarded those parameters to 
which neither {\Ca} and \mbox{RhoGTP} are sensitive 
by the quantitative criterion indicated above.  Then in Table~\ref{tab:tableVII} we 
cross-listed those parameters 
to which {\Ca} or \mbox{RhoGTP} or both were sensitive. 
According to the adopted criterion of importance, 
our analysis shows that $k_7$ and $k_8$ are the most important parameters 
to reproduce the expected system behavior, because they respectively have ten and six entries 
in Table~\ref{tab:tableVII}.
In Figs.~\ref{fig:k7_ca}~and~\ref{fig:k7_rho} ,as an example, we show the cumulative distribution functions~(\ref{eq:cumul}) calculated by
Monte Carlo filtering for the parameter $k_7$ in the case $y_i=\mathrm{Ca^{2+}}$ and 
$y_i=\mathrm{RhoGTP}$ respectively. The red curves represent 
the distribution of ``acceptable'' parameters,
whereas the blue ones represent the distributions of ``unacceptable'' parameters.
In panel~A we report the results relative to the sensitivity test performed with 
the objective function calculated as in~(\ref{eq:gbj}), 
panel~B shows the results of the Monte Carlo filtering method
with the objective function estimated according to~(\ref{eq:Gbj}), and panel~C~and~D 
are relative to the choice of the objective functions in~(\ref{eq:obj-pk}) respectively.
Similarly, In Fig.~\ref{fig:k26_rho},  we show the cumulative distribution functions~(\ref{eq:cumul}) calculated by
Monte Carlo filtering for the parameter $k_{26}$ in the case $y_i=\mathrm{RhoGTP}$.

Examination of Tables~\ref{tab:tableI}--\ref{tab:tableVII} permits one to classify 
the input parameters into three broad categories:
\subsubsection*{RhoGTP-Only Sensitivity Parameters (four or more entries in Table~\ref{tab:tableVII})}
\begin{description}
\item{$\text{k}_{26}$\ } PAR1 to G13 binding rate;
\item{$\text{k}_{36}$\ } G13GTP-GEF dissociation rate;
\item{$\text{k}_{37}$\ } GEF hydrolyzing G13GTP;
\item{$\text{k}_{40}$\ } $\mathrm{GEFG13^*}$ to Rho binding rate;
\item{$\text{k}_{43}$\ } Rho deactivation rate.
\end{description}
Variations of these parameter affect only \mbox{RhoGTP} 
and its functionals
\begin{equation}\label{eq:rho_func}
\begin{array}{ll}
\text{[RhoGTP]}(t_i,\bl{k}), \quad
&\int_0^T\text{[RhoGTP]}(t,\bl{k})dt\\
{}\\
\text{[RhoGTP]}^{\text{max}}(\bl{k}), \quad&\tpkrho(\bl{k})
\end{array}
\end{equation}
introduced in (\ref{eq:states})--(\ref{eq:int_states}), and 
(\ref{eq:value-pk}). Neither small variations (Taylor's expansion), 
nor large variations (Monte-Carlo filtering) of these 
parameters about their nominal values  affect {\Ca} and 
its functionals. This is an expected result as these parameters 
only belong to the \mbox{RhoGTP} module downstream 
of $\mathrm{G13}$ (see Fig.~\ref{fig:path}).

\input tableVII.tex
Interestingly, our sensitivity analysis found, 
independently of  \cite{mclaughlin_2005}, that 
$\text{k}_{26}$, which represents PAR1 binding 
rate to $\text{G}_{12/13}$, 
is a key factor in this model to reproduce the experimental 
data. This emerges from two different sensitivity 
analysis methods and constitutes  a further 
test of validity for the signaling model 
adopted in \cite{mclaughlin_2005}.
\subsubsection*{\Ca-Only Sensitivity Parameters (four or more entries in Table~\ref{tab:tableVII})}
\begin{description}
\item{$\text{k}_{19}$\ } rate of hydrolysis of GqGTP by PLC$\be$;
\item{$\text{k}_{22}$\ } PLC$\be$-PIP2 binding rate;
\item{$\text{k}_{25}$\ } IP3 rate of consumption;
\end{description}
Variations of these parameter affect only {\Ca} 
and its functionals
\begin{equation}\label{eq:ca_func}
\begin{array}{ll}
\tCa(t_i,\bl{k}), \quad
&\int_0^T\tCa(t,\bl{k})dt\\
{}\\
\tCa^{\text{max}}(\bl{k}), \quad&\tpkca(\bl{k})
\end{array}
\end{equation}
introduced in (\ref{eq:states})--(\ref{eq:int_states}), and 
(\ref{eq:value-pk}). Small variations (Taylor's expansion) 
or large variations (Monte-Carlo filtering) of these 
parameters about their nominal values do not affect 
\mbox{RhoGTP} or its functionals. Again, this is an expected result as 
these parameters  only belong to the 
$\mathrm{Ca^{2+}}$ signaling module downstream Gq.

The presence of parameters that affect  the $\mathrm{Ca^{2+}}$
output but not the RhoGTP output and viceversa is due to the 
modular structure of the model in~\cite{mclaughlin_2005}.
In this model, Gq and G13 pathways are described by two separate computational modules with a  
common  input, i.e. the activated PAR1 (Fig.~\ref{fig:path}).

The {\Ca} functionals in (\ref{eq:ca_func}) are not sensitive
to small variations of $\text{k}_{25}$ (rate of IP3 consumption) 
about its nominal value $\bar{\text{k}}_{25}$ (Taylor's method 
for $|\Delta \text{k}_{25}|\le 5\%$). They are however severely 
sensitive for variations of $\text{k}_{25}$ in the range 
$(10^{-1}\bar{\text{k}}_{25},10\bar{\text{k}}_{25})$ (Monte 
Carlo filtering method). 

In all cases however, the notion of 
``sensitivity'' and ``relevance'' is the same by both methods, 
pointing to a robustness and self-consistency 
of the model in \cite{mclaughlin_2005}.

\subsubsection*{\Ca-RhoGTP Sensitivity Parameters}
\begin{description}
\item{$\text{k}_{7}$\ } PAR1 deactivation rate;
\item{$\text{k}_{8}$\ } PAR1-Gq binding rate;
\item{$\text{k}_{14}$\ } rate of release of $\beta\gamma$ by Gq;
\end{description}
PAR1 acts on the common part of the {\Ca} and \mbox{RhoGTP} 
pathways and accordingly both of them and their functionals 
are sensitive to both these parameters. 
While PAR1 deactivation rate $\text{k}_7$ is important 
to both {\Ca} and \mbox{RhoGTP}, the various components of 
the system respond differently to variations of this parameter. 
First, {\Ca} and its functionals (\ref{eq:ca_func}), are not sensitive to small 
variations of $\text{k}_7$ about its nominal value $\bar{\text{k}}_7$ 
(Taylor's methods for $|\Delta\text{k}_7|\le 5\%$). On the other 
hand all the {\Ca} functionals are very sensitive for variations 
of $\text{k}_7$ in the range $(10^{-1}\bar{\text{k}}_7,10\bar{\text{k}}_7)$ 
(Monte Carlo filtering method).

Contrarily \mbox{RhoGTP} and its functionals are very sensitive 
to any variation $\text{k}_7$, small or large, and by whatever 
method sensitivity is evaluated  (Taylor or Monte Carlo filtering).

Finally, we observe that the pathways analyzed in 
this work are shared in a number of signaling cellular systems. 
In a recent work, we devised a mathematical model of PAR1-mediated signaling in human platelets~\cite{lenoci_2010}
that shares several features with the model in~\cite{mclaughlin_2005}. For these reasons and for 
the generality of the methods adopted in this work, the conclusions of this  study can,
at least qualitatively, be 
extended to the model in~\cite{lenoci_2010}.

\end{document}

%% file: tableI.tex
\begin{table}[!ht]
\caption{
\bf{The sensitivity coefficients calculated by the Taylor-like formula at the peak time $\tpk$ for 
$\mathrm{Ca^{2+}}$ and $\mathrm{RhoGTP}$}}
\scriptsize
\centering
\begin{tabular}{|l|l|l|l|l|p{2in}|}
\hline
      & \multicolumn{2}{|c|}{$\mathbf{\mathrm{Ca^{2+}}}$} & \multicolumn{2}{|c|}{$\mathbf{\mathrm{RhoGTP}}$}&   \\
\hline
{\bf k} &        $\mathbf{S_{+}}$ &        $\mathbf{S_{-}}$&          $\mathbf{S_{+}}$      &          $\mathbf{S_{-}}$       &  {\bf FUNCTIONAL ROLE}   \\
\hline
1	&	  0.000	&	  0.002	&	   0.012	&	   0.038	&	 PAR1 activation by SFLLRN\\
\hline
2	&	  0.000	&	  0.000	&	   0.000	&	   0.000	&	\\
\hline
3	&	 -0.001	&	  0.001	&	  -0.004	&	  -0.018	&	\\
\hline
\bf 7	&	 -0.006	&	 -0.006	&\bf	  -0.193	&\bf	  -0.197	&	PAR1 deactivation\\
\hline
8	&	  0.006	&	  0.006	&	  -0.051	&	  -0.050	&	 PAR1 binding Gq\\
\hline
9	&	 -0.001	&	 -0.001	&	   0.008	&	   0.009	&	\\
\hline
10	&	  0.008	&	  0.010	&	   0.003	&	  -0.007	&	Gq releasing GDP\\
\hline
11	&	  0.000	&	  0.000	&	   0.000	&	   0.000	&	\\
\hline
12	&	  0.000	&	  0.000	&	   0.001	&	   0.001	&	 Gq binding GTP\\
\hline
13	&	  0.000	&	  0.000	&	   0.000	&	   0.000	&	\\
\hline
14	&	  0.019	&	  0.024	&	   0.056	&	   0.053	&	Gq releasing $\beta\gamma$\\
\hline
15	&	  0.000	&	  0.000	&	   0.000	&	   0.000	&	 \\
\hline
16	&	  0.000	&	  0.000	&	   0.000	&	   0.000	&	Gq deactivation\\
\hline
17	&	  0.000	&	  0.000	&	   0.014	&	   0.000	&	${\rm PLC\beta}$ binding GqGTP\\
\hline
18	&	  0.000	&	  0.000	&	   0.000	&	   0.000	&	 \\
\hline
19	&	 -0.040	&	 -0.034	&	  -0.018	&	  -0.003	&	${\rm PLC\beta}$ hydrolizing GqGTP\\
\hline
20	&	  0.000	&	  0.000	&	   0.000	&	   0.000	&	${\rm PLC\beta}$ releasing GqGDP\\
\hline
21	&	  0.000	&	  0.000	&	   0.000	&	   0.000	&	\\
\hline
22	&	  0.032	&	  0.042	&	   0.007	&	   0.013	&	${\rm PLC\beta}$ binding PIP2\\
\hline
23	&	  0.000	&	  0.000	&	   0.000	&	   0.000	&	\\
\hline
24	&	  0.000	&	  0.000	&	   0.000	&	   0.000	&	${\rm PLC\beta}$ hydrolizing PIP2\\
\hline
25	&	 -0.065	&	 -0.066	&	  -0.016	&	  -0.009	&	 Consumption of IP3\\
\hline
\bf 26	&	  0.000	&	  0.000	&\bf	   0.245	&\bf	   0.247	&	PAR1 binding G13\\
\hline
27	&	  0.000	&	  0.000	&	  -0.041	&	  -0.041	&	\\
\hline
28	&	  0.000	&	  0.000	&	   0.039	&	   0.042	&	G13 releasing GDP\\
\hline
29	&	  0.000	&	  0.000	&	   0.000	&	   0.000	&	\\
\hline
30	&	  0.000	&	  0.000	&	   0.000	&	   0.000	&	G13 binding GTP\\
\hline
31	&	  0.000	&	  0.000	&	   0.000	&	   0.000	&	\\
\hline
32	&	  0.000	&	  0.000	&	  -0.001	&	  -0.001	&	G13 releasing $\beta\gamma$\\
\hline
33	&	  0.000	&	  0.000	&	   0.000	&	   0.000	&	\\
\hline
34	&	  0.000	&	  0.000	&	  -0.022	&	  -0.022	&	G13 deactivation\\
\hline
\bf 35	&	  0.000	&	  0.000	&\bf	   0.106	&\bf	   0.112	&	 GEF binding G13GTP\\
\hline
\bf 36	&	  0.000	&	  0.000	&\bf	  -0.107	&\bf	  -0.111	&	\\
\hline
37	&	  0.000	&	  0.000	&	  -0.054	&	  -0.054	&	GEF hydrolizing G13GTP\\
\hline
38	&	  0.000	&	  0.000	&	   0.000	&	   0.000	&	GEF releasing G13GDP\\
\hline
39	&	  0.000	&	  0.000	&	   0.000	&	   0.000	&	\\
\hline
\bf 40	&	  0.000	&	  0.000	&\bf	   0.249	&\bf	   0.251	&	 GEFG13GTP binding RhoGDP\\
\hline
41	&	  0.000	&	  0.000	&	  -0.017	&	  -0.017	&	\\
\hline
42	&	  0.000	&	  0.000	&	   0.017	&	   0.018	&	Rho activation\\
\hline
\bf 43	&	  0.000	&	  0.000	&\bf	  -0.199	&\bf	  -0.203	&	  Rho deactivation\\
\hline
\end{tabular}
\normalsize
\begin{flushleft}
The sensitiviy coefficients $S_{ij}(t,\bbk)=\frac{\partial\ln(y_i)}{\partial\ln(k_j)}
\Bigg|_{\bl{k}=\bbk}$ calculated at the nominal peak time $\tpk$ for $y_i=\mathrm{Ca^{2+}}$ and variations $\mathbf{\Delta k}=\pm 5\%$
(columns two and three) 
and $y_i=\mathrm{RhoGTP}$ and variations $\mathbf{\Delta k}=\pm 5\%$
(columns four and five). 
Sensitivity coefficients  larger than $10\%$ are in bold font.
The nominal peak time $\tpk$ is defined in (\ref{eq:time-pk}) and 
the coefficients $S_{ij}$ are defined in (\ref{eq:matrix_s}). 
\end{flushleft}
\label{tab:tableI}
\end{table}

%% file: tableII.tex
\begin{table}[!ht]
\caption{
\bf{The sensitivity coefficients calculated by the Taylor-like formula over the course of the entire simulation 
$T=600~\mathrm{s}$ for $\mathrm{Ca^{2+}}$ and $\mathrm{RhoGTP}$}}
\centering
\scriptsize
\begin{tabular}{|l|l|l|l|l|p{2in}|}
\hline
    & \multicolumn{2}{|c|}{$\mathbf{\mathrm{Ca^{2+}}}$} & \multicolumn{2}{|c|}{$\mathbf{\mathrm{RhoGTP}}$}& \\
\hline
{\bf k}        &        $\mathbf{\Sigma_{+}}$ &        $\mathbf{\Sigma_{-}}$&          $\mathbf{\Sigma_{+}}$      &          $\mathbf{\Sigma_{-}}$       &{\bf FUNCTIONAL ROLE}     \\
\hline
1	&	  0.000	&	  0.000	&	   0.008	&	  -0.017	&	 PAR1 activation by SFLLRN\\
\hline
2	&	  0.000	&	  0.000	&	   0.000	&	   0.000	&	\\
\hline
3	&	  0.000	&	  0.000	&	  -0.003	&	  -0.013	&	\\
\hline
\bf 7	&	  0.000	&	  0.000	&\bf	  -0.449	&\bf	  -0.443	&	PAR1 deactivation\\
\hline
8	&	  0.000	&	  0.000	&	   0.021	&	   0.040	&	 PAR1 binding Gq\\
\hline
9	&	  0.000	&	  0.000	&	  -0.007	&	  -0.007	&	\\
\hline
10	&	  0.000	&	  0.000	&	  -0.001	&	  -0.012	&	Gq releasing GDP\\
\hline
11	&	  0.000	&	  0.000	&	   0.000	&	   0.000	&	\\
\hline
12	&	  0.000	&	  0.000	&	  -0.001	&	  -0.001	&	 Gq binding GTP\\
\hline
13	&	  0.000	&	  0.000	&	   0.000	&	   0.000	&	\\
\hline
14	&	  0.000	&	  0.000	&	  -0.006	&	  -0.016	&	Gq releasing $\beta\gamma$\\
\hline
15	&	  0.000	&	  0.000	&	   0.000	&	   0.000	&	 \\
\hline
16	&	  0.000	&	  0.000	&	   0.000	&	   0.000	&	Gq deactivation\\
\hline
17	&	  0.000	&	  0.000	&	   0.011	&	   0.000	&	${\rm PLC\beta}$ binding GqGTP\\
\hline
18	&	  0.000	&	  0.000	&	   0.000	&	   0.000	&	 \\
\hline
19	&	  0.000	&	  0.000	&	  -0.022	&	   0.002	&	${\rm PLC\beta}$ hydrolizing GqGTP\\
\hline
20	&	  0.000	&	  0.000	&	   0.000	&	   0.000	&	${\rm PLC\beta}$ releasing GqGDP\\
\hline
21	&	  0.000	&	  0.000	&	   0.000	&	   0.000	&	\\
\hline
22	&	  0.000	&	  0.000	&	   0.001	&	   0.019	&	${\rm PLC\beta}$ binding PIP2\\
\hline
23	&	  0.000	&	  0.000	&	   0.000	&	   0.000	&	\\
\hline
24	&	  0.000	&	  0.000	&	   0.000	&	   0.000	&	${\rm PLC\beta}$ hydrolizing PIP2\\
\hline
25	&	 -0.051	&	 -0.057	&	  -0.009	&	   0.031	&	 Consumption of IP3\\
\hline
\bf 26	&	  0.000	&	  0.000	&\bf 	   0.423	&\bf	   0.429	&	PAR1 binding G13\\
\hline
27	&	  0.000	&	  0.000	&	  -0.070	&	  -0.071	&	\\
\hline
28	&	  0.000	&	  0.000	&	   0.068	&	   0.074	&	G13 releasing GDP\\
\hline
29	&	  0.000	&	  0.000	&	   0.000	&	   0.000	&	\\
\hline
30	&	  0.000	&	  0.000	&	   0.000	&	   0.000	&	G13 binding GTP\\
\hline
31	&	  0.000	&	  0.000	&	   0.000	&	   0.000	&	\\
\hline
32	&	  0.000	&	  0.000	&	   0.000	&	   0.000	&	G13 releasing $\beta\gamma$\\
\hline
33	&	  0.000	&	  0.000	&	   0.000	&	   0.000	&	\\
\hline
\bf 34	&	  0.000	&	  0.000	&\bf 	  -0.199	&\bf	  -0.202	&	G13 deactivation\\
\hline
35	&	  0.000	&	  0.000	&	   0.064	&	   0.069	&	 GEF binding G13GTP\\
\hline
36	&	  0.000	&	  0.000	&	  -0.066	&	  -0.066	&	\\
\hline
\bf 37	&	  0.000	&	  0.000	&\bf	  -0.488	&\bf	  -0.504	&	GEF hydrolizing G13GTP\\
\hline
38	&	  0.000	&	  0.000	&	   0.000	&	   0.000	&	GEF releasing G13GDP\\
\hline
39	&	  0.000	&	  0.000	&	   0.000	&	   0.000	&	\\
\hline
\bf 40	&	  0.000	&	  0.000	&\bf	   0.429	&\bf	   0.434	&	 GEFG13GTP binding RhoGDP\\
\hline
41	&	  0.000	&	  0.000	&	  -0.030	&	  -0.030	&	\\
\hline
42	&	  0.000	&	  0.000	&	   0.029	&	   0.031	&	Rho activation\\
\hline
\bf 43	&	  0.000	&	  0.000	&\bf	  -0.402	&\bf	  -0.421	&	  Rho deactivation\\
\hline
\end{tabular}
\normalsize
\begin{flushleft}
The sensitiviy coefficients $\Sig_{ij}(\bbk)=\frac{\bar{k}_j}{z_i(\bbk)}
\frac{\partial z_i}{\partial k_j}
(\bl{k})\Biggm|_{\bl{k}=\bar{\bl{k}}}$ with $\bl{z}(\bl{k})=\int_0^T \bl{y}(t,\bl{k})dt$ for $y_i=\mathrm{Ca^{2+}}$ and variations $\mathbf{\Delta k}=\pm 5\%$
(columns two and three) 
and $y_i=\mathrm{RhoGTP}$ and variations $\mathbf{\Delta k}=\pm 5\%$
(columns four and five). Sensitivity coefficients  larger than $10\%$ are in bold font.
\end{flushleft}
\label{tab:tableII}
\end{table}

%% file: tableVIII.tex
\begin{sidewaystable}
\centering
\caption{
\bf{Symbolic reaction schemes and effective kinetic parameters used in the
simulations of  the PAR1-mediated activation model of endothelial cells proposed in~\cite{mclaughlin_2005}.}}
\begin{scriptsize}
\begin{tabular}{|p{3.4in}|p{1.2in}|p{1.2in}|p{0.6in}|}
\hline
{\bf Symbolic Reactions} & $\mathbf{{\rm k_{\rightarrow}}}$ & $\mathbf{{\rm k_{\leftarrow}}}$ & {\bf References}\\
\hline
\multicolumn{4}{|c|}{\emph{Reactions governing PAR1 activation}} \\
\hline
${\rm Agonist + PAR1 \rightleftharpoons  PAR1^*}$ &  6.00E+04 ${\rm M^{-1}s^{-1}}$ $(k_1)$&  1.00E-03 ${\rm s^{-1}}$$(k_2)$  
& \cite{ahn_1997}\\
\hline
${\rm Agonist  \rightarrow  null}$ & 2.00E-01 ${\rm s^{-1}}$ $(k_3)$&&\cite{mclaughlin_2005}\\
\hline
${\rm PAR1^*  \rightarrow  null}$ & 2.00E+01 ${\rm s^{-1}}$ $(k_7)$& &\cite{mclaughlin_2005}\\
\hline
\multicolumn{4}{|c|}{\emph{Reactions governing Gq activation}} \\
\hline
${\rm PAR1^*+G_qGDP\cdot \beta\gamma \rightleftharpoons  PAR1^*\cdot G_qGDP\cdot \beta\gamma}$ 
& 1.00E+08 ${\rm M^{-1}s^{-1}}$ $(k_8)$ &1.00 ${\rm s^{-1}}$ $(k_9)$& \cite{zhong_2003}\\
\hline
${\rm  PAR1^*\cdot G_qGDP\cdot \beta\gamma \rightleftharpoons PAR1^*\cdot G_q\cdot \beta\gamma + GDP}$ 
& 5.00 ${\rm s^{-1}}$ $(k_{10})$& 1.00E+06 ${\rm M^{-1}s^{-1}}$ $(k_{11})$& \cite{zhong_2003}\\
\hline
${\rm PAR1^*\cdot G_q\cdot \beta\gamma + GTP \rightleftharpoons PAR1^*\cdot G_qGTP\cdot \beta\gamma}$
&1.00E+06 ${\rm M^{-1}s^{-1}}$ $(k_{12})$& 1.00E-01 ${\rm s^{-1}}$ $(k_{13})$&\cite{zhong_2003}\\
\hline
${\rm PAR1^*\cdot G_qGTP\cdot \beta\gamma \rightleftharpoons PAR1^* + G_qGTP +\beta\gamma}$
& 2.00 ${\rm s^{-1}}$ $(k_{14})$& 1.00E+07 ${\rm M^{-2}s^{-1}}$ $(k_{15})$&\cite{zhong_2003}\\  
\hline
${\rm  G_qGTP \rightarrow G_qGDP}$ & 2.00E-02 ${\rm s^{-1}}$ $(k_{16})$& & \cite{biddlecome_1996,ross_1999,ross_2000}\\
\hline
\multicolumn{4}{|c|}{\emph{Reactions governing IP3 and DAG generation}} \\
\hline
${\rm PLC\beta + G_qGTP \rightleftharpoons PLC\beta\cdot G_qGTP}$ &
5.00E+08 ${\rm M^{-1}s^{-1}}$ $(k_{17})$& 5.00 ${\rm s^{-1}}$ $(k_{18})$& \cite{ross_2000,blank_1992}\\
\hline
${\rm PLC\beta\cdot G_qGTP  \rightarrow PLC\beta\cdot G_qGDP}$ & 1.5E+01 ${\rm s^{-1}}$ $(k_{19})$& 
& \cite{ross_1999,ross_2000}\\
\hline
${\rm PLC\beta\cdot G_qGDP \rightleftharpoons PLC\beta + G_qGDP}$
& 1.00E+05 ${\rm s^{-1}}$ $(k_{20})$&  1.00E+02 ${\rm M^{-1}s^{-1}}$ $(k_{21})$& \cite{lemon_2003}\\
\hline
${\rm PLC\beta\cdot G_qGTP + PIP2 \rightleftharpoons  PLC\beta\cdot G_qGTP\cdot PIP2}$
& 1.00E+09 ${\rm M^{-1}s^{-1}}$ $(k_{22})$& 1.00 ${\rm s^{-1}}$ $(k_{23})$& \cite{lemon_2003}\\
\hline
${\rm PLC\beta\cdot G_qGTP\cdot PIP2 \rightarrow PLC\beta\cdot G_qGTP + IP3/DAG}$
&  1.00E+02 ${\rm s^{-1}}$ $(k_{24})$& & \cite{lemon_2003}\\
\hline
${\rm IP3\rightarrow  null}$ &  2.40E-02 ${\rm s^{-1}}$ $(k_{25})$& & \cite{mclaughlin_2005}\\
\hline
\multicolumn{4}{|c|}{\emph{Reactions governing Ca2+ mobilization}}  \\
\hline
As described previously & & &\cite{keizer_1992}\\
\hline
\multicolumn{4}{|c|}{\emph{Reactions governing G12/13 activation}} \\ 
\hline
${\rm PAR1^*+G_{12/13}GDP\cdot \beta\gamma \rightleftharpoons  PAR1^*\cdot G_{12/13}GDP\cdot \beta\gamma}$ 
& 1.00E+08 ${\rm M^{-1}s^{-1}}$  $(k_{26})$&1.00 ${\rm s^{-1}}$ $(k_{27})$& \cite{zhong_2003}\\
\hline
${\rm  PAR1^*\cdot G_{12/13}GDP\cdot \beta\gamma \rightleftharpoons PAR1^*\cdot G_{12/13}\cdot \beta\gamma + GDP}$ 
& 6.00 ${\rm s^{-1}}$ $(k_{28})$& 1.00E+06 ${\rm M^{-1}s^{-1}}$ $(k_{29})$& \cite{zhong_2003}\\
\hline
${\rm PAR1^*\cdot G_{12/13}\cdot \beta\gamma + GTP \rightleftharpoons PAR1^*\cdot G_{12/13}GTP\cdot \beta\gamma}$
&1.00E+06 ${\rm M^{-1}s^{-1}}$ $(k_{30})$& 1.00E-01 ${\rm s^{-1}}$ $(k_{31})$&\cite{zhong_2003}\\
\hline
${\rm PAR1^*\cdot G_{12/13}GTP\cdot \beta\gamma \rightleftharpoons PAR1^* + G_{12/13}GTP +\beta\gamma}$
& 2.00 ${\rm s^{-1}}$ $(k_{32})$& 1.00E+07 ${\rm M^{-2}s^{-1}}$ $(k_{33})$&\cite{zhong_2003}\\  
\hline
${\rm  G_{12/13}GTP \rightarrow G_{12/13}GDP}$ & 4.00E-03 ${\rm s^{-1}}$ $(k_{34})$& & \cite{kozasa_1998,singer_1994}\\
\hline
\multicolumn{4}{|c|}{\emph{Reactions governing Rho activation}} \\
\hline
${\rm GEF_{Rho}+ G_{12/13}GTP  \rightleftharpoons GEF_{Rho}\cdot G_{12/13}GTP}$
& 1.00E+09 ${\rm M^{-1}s^{-1}}$ $(k_{35})$& 3.00 ${\rm s^{-1}}$ $(k_{36})$& \cite{ross_2000}\\
\hline
${\rm GEF_{Rho}\cdot G_{12/13}GTP \rightarrow  GEF_{Rho}\cdot G_{12/13}GDP}$
& 1.00E-2 ${\rm s^{-1}}$ $(k_{36})$& & \cite{kozasa_1998}\\
\hline
${\rm GEF_{Rho}\cdot G_{12/13}GDP \rightleftharpoons GEF_{Rho} + G_{12/13}GDP}$
&1.00E+06 ${\rm s^{-1}}$ $(k_{38})$& 3.00 ${\rm M^{-1}s^{-1}}$ $(k_{39})$& \cite{mclaughlin_2005}\\
\hline
${\rm GEF_{Rho}\cdot G_{12/13}GTP + RhoGDP \rightleftharpoons}$ ${\rm GEF_{Rho}\cdot G_{12/13}GTP\cdot RhoGDP}$
& 7.70E+09 ${\rm M^{-1}s^{-1}}$ $(k_{40})$&  7.70E+02 ${\rm s^{-1}}$ $(k_{41})$& \cite{mclaughlin_2005}\\
\hline
${\rm GEF_{Rho}\cdot G_{12/13}GTP\cdot RhoGDP \rightarrow}$ ${\rm GEF_{Rho}\cdot G_{12/13}GTP + RhoGTP}$
& 1.03E+04 ${\rm s^{-1}}$  $(k_{42})$& & \cite{mclaughlin_2005}\\
\hline
${\rm RhoGTP \rightarrow RhoGDP}$ & 6.00E-02 ${\rm s^{-1}}$ $(k_{43})$& & \cite{yamamoto_1988}\\
\hline
\end{tabular}
\end{scriptsize}
\begin{flushleft}
The nominal values of the parameters adopted in the PAR1-mediated activation
model of endothelial cells proposed in~\cite{mclaughlin_2005}. The kinetic constants are 
numbered sequentially starting from the first $k_{\rightarrow}$ ($k_1$) and proceeding in reading order with
the exception of reaction ``${\rm PAR1^*  \rightarrow  null}$'' which is regulated by the constant $k_7$. 
\end{flushleft}
\label{tab:tableVIII}
\end{sidewaystable}

%% file: tableIII.tex
\begin{table}[!ht]
\caption{
\bf{The sensitivity coefficients calculated by  Monte Carlo filtering at the nominal peak time $\tpk$ for 
$\mathrm{Ca^{2+}}$ and $\mathrm{RhoGTP}$}}
\centering
\scriptsize
\begin{tabular}{|l|l|l|p{2in}|}
\hline
        &   \multicolumn{2}{|c|}{$\mathbf{\mathrm{d_{1,2}}}$} &\\
\hline
{\bf k}       & $\mathbf{\mathrm{Ca^{2+}}}$ & $\mathbf{\mathrm{RhoGTP}}$& {\bf FUNCTIONAL ROLE}  \\
\hline
1	&	  0.095	&	  0.084	&	 PAR1 activation by SFLLRN\\
\hline
2	&	  0.031	&	  0.026	&	\\
\hline
\bf 3	&\bf	  0.113	&	  0.093	&	\\
\hline
\bf 7	&\bf	  0.160	&\bf	  0.285	&	PAR1 deactivation\\
\hline
\bf 8	&	  0.091	&\bf	  0.153	&	 PAR1 binding Gq\\
\hline
9	&	  0.037	&	  0.043	&	\\
\hline
10	&	 0.072	&	 0.017	&	Gq releasing GDP\\
\hline
11	&	 0.017	&	 0.022	&	\\
\hline
12	&	 0.018	&	 0.021	&	 Gq binding GTP\\
\hline
13	&	 0.041	&	 0.019	&	\\
\hline
\bf 14	&\bf	 0.153	&	 0.061	&	Gq releasing $\beta\gamma$\\
\hline
15	&	 0.023	&	 0.012	&	 \\
\hline
16	&	 0.026	&	 0.024	&	Gq deactivation\\
\hline
17	&	 0.024	&	 0.033	&	${\rm PLC\beta}$ binding GqGTP\\
\hline
18	&	 0.030	&	 0.038	&	 \\
\hline
\bf 19	&\bf	 0.421	&	 0.034	&	${\rm PLC\beta}$ hydrolizing GqGTP\\
\hline
20	&	 0.042	&	 0.013	&	${\rm PLC\beta}$ releasing GqGDP\\
\hline
21	&	 0.029	&	 0.029	&	\\
\hline
\bf 22	&\bf	 0.394	&	 0.017	&	${\rm PLC\beta}$ binding PIP2\\
\hline
23	&	 0.031	&	 0.030	&	\\
\hline
24	&	 0.034	&	 0.021	&	${\rm PLC\beta}$ hydrolizing PIP2\\
\hline
\bf 25	&\bf	 0.413	&	 0.018	&	 Consumption of IP3\\
\hline
\bf 26	&	 0.035	&\bf	 0.391	&	PAR1 binding G13\\
\hline
\bf 27	&	 0.016	&\bf	 0.105	&	\\
\hline
28	&	 0.027	&	 0.068	&	G13 releasing GDP\\
\hline
29	&	 0.018	&	 0.026	&	\\
\hline
30	&	 0.042	&	 0.027	&	G13 binding GTP\\
\hline
31	&	 0.017	&	 0.015	&	\\
\hline
32	&	 0.038	&	 0.019	&	G13 releasing $\beta\gamma$\\
\hline
33	&	 0.023	&	 0.048	&	\\
\hline
34	&	 0.017	&	 0.073	&	G13 deactivation\\
\hline
\bf 35	&	 0.021	&\bf	 0.158	&	 GEF binding G13GTP\\
\hline
\bf 36	&	 0.024	&\bf	 0.196	&	\\
\hline
\bf 37	&	 0.026	&\bf	 0.170	&	GEF hydrolizing G13GTP\\
\hline
38	&	 0.016	&	 0.043	&	GEF releasing G13GDP\\
\hline
39	&	 0.026	&	 0.025	&	\\
\hline
\bf 40	&	 0.042	&\bf	 0.385	&	 GEFG13GTP binding RhoGDP\\
\hline
\bf 41	&	 0.024	&\bf	 0.104	&	\\
\hline
42	&	 0.029	&	 0.031	&	Rho activation\\
\hline
\bf 43	&	 0.023	&\bf	 0.350	&	  Rho deactivation\\
\hline
\end{tabular}
\normalsize
\begin{flushleft}
The sensitivity coefficients measured as Kolmogorov-Smirnov distances $d_{1,2}$ estimated in the ranges $[1/10,10]\overline k_j$
at the nominal peak time  $\tpk$ for $y_i=\mathrm{Ca^{2+}}$ (column two) and $y_i=\mathrm{RhoGTP}$ (column three).
Sensitivity coefficients  larger than $10\%$ are in bold font.
The nominal peak time $\tpk$ is defined in (\ref{eq:time-pk}) and the objective function 
used in the simulations is defined in~(\ref{eq:gbj}).
\end{flushleft}
\label{tab:tableIII}
\end{table}

%% file: tableIV.tex
\begin{table}[!ht]
\caption{
\bf{The sensitivity coefficients calculated by Monte Carlo filtering over the course of the entire simulation $T=600~\mathrm{s}$ for $\mathrm{Ca^{2+}}$ and $\mathrm{RhoGTP}$}}
\centering
\scriptsize
\begin{tabular}{|l|l|l|p{2in}|}
\hline
        &   \multicolumn{2}{|c|}{$\mathbf{\mathrm{d_{1,2}}}$} &\\
\hline
{\bf k}       & $\mathbf{\mathrm{Ca^{2+}}}$ & $\mathbf{\mathrm{RhoGTP}}$& {\bf FUNCTIONAL ROLE}  \\
\hline
1	&0.042	 	&	  0.032	&	 PAR1 activation by SFLLRN\\
\hline
2	&0.034	 	&	  0.029	&	\\
\hline
3	&0.045	 	&	  0.071	&	\\
\hline
\bf 7	&\bf 0.171	 	&\bf	  0.208	&	PAR1 deactivation\\
\hline
\bf 8	&\bf 0.109	 	&	  0.049	&	 PAR1 binding Gq\\
\hline
9	&0.041	 	&	  0.017	&	\\
\hline
10	&0.044	 	&	 0.017	&	Gq releasing GDP\\
\hline
11	&0.017	 	&	 0.018	&	\\
\hline
12	&0.046	 	&	 0.021	&	 Gq binding GTP\\
\hline
13	&0.020	 	&	 0.029	&	\\
\hline
14	&0.034	 	&	 0.023	&	Gq releasing $\beta\gamma$\\
\hline
15	&0.030	 	&	 0.019	&	 \\
\hline
16	&0.031	 	&	 0.038	&	Gq deactivation\\
\hline
17	&0.021	 	&	 0.050	&	${\rm PLC\beta}$ binding GqGTP\\
\hline
18	&0.020	 	&	 0.028	&	 \\
\hline
\bf 19	&\bf 0.157	 	&	 0.022	&	${\rm PLC\beta}$ hydrolizing GqGTP\\
\hline
20	&0.035	 	&	 0.030	&	${\rm PLC\beta}$ releasing GqGDP\\
\hline
21	&0.042	 	&	 0.011	&	\\
\hline
22	&0.085	 	&	 0.020	&	${\rm PLC\beta}$ binding PIP2\\
\hline
23	&0.020	 	&	 0.013	&	\\
\hline
24	&0.028	 	&	 0.018	&	${\rm PLC\beta}$ hydrolizing PIP2\\
\hline
\bf 25	&\bf 0.216	 	&	 0.043	&	 Consumption of IP3\\
\hline
\bf 26	&0.032	 	&\bf	 0.141	&	PAR1 binding G13\\
\hline
27	&0.034	 	&	 0.041	&	\\
\hline
28	&0.026	 	&	 0.044	&	G13 releasing GDP\\
\hline
29	&0.031	 	&	 0.048	&	\\
\hline
30	&0.033	 	&	 0.028	&	G13 binding GTP\\
\hline
31	&0.022	 	&	 0.024	&	\\
\hline
32	&0.050	 	&	 0.021	&	G13 releasing $\beta\gamma$\\
\hline
33	&0.029	 	&	 0.017	&	\\
\hline
34	&0.038	 	&	 0.066	&	G13 deactivation\\
\hline
35	&0.034	 	&	 0.058	&	 GEF binding G13GTP\\
\hline
\bf 36	&0.029	 	&\bf 	 0.111	&	\\
\hline
\bf 37	&0.033	 	&\bf	 0.125	&	GEF hydrolizing G13GTP\\
\hline
38	&0.058	 	&	 0.011	&	GEF releasing G13GDP\\
\hline
39	&0.033	 	&	 0.016	&	\\
\hline
\bf 40	&0.044	 	&\bf	 0.105	&	 GEFG13GTP binding RhoGDP\\
\hline
41	&0.045	 	&	 0.052	&	\\
\hline
42	&0.052	 	&	 0.039	&	Rho activation\\
\hline
\bf 43	&0.037	 	&\bf	 0.238	&	  Rho deactivation\\
\hline
\end{tabular}
\normalsize
\begin{flushleft}
The sensitivity coefficients measured as Kolmogorov-Smirnov distances $d_{1,2}$  estimated in the ranges $[1/10,10]~\overline k_j$ over the
course of the entire simulation $T=600~\mathrm{s}$ for $y_i=\mathrm{Ca^{2+}}$ (column two) and $y_i=\mathrm{RhoGTP}$ (column three).
Sensitivity coefficients larger than $10\%$ are in bold font.
The objective function used in the simulations is defined in (\ref{eq:Gbj}).
\end{flushleft}
\label{tab:tableIV}
\end{table}

%% file: tableV.tex
\begin{table}[!ht]
\caption{
\bf{The sensitivity coefficients calculated by Monte Carlo filtering for the peak values of $\mathrm{Ca^{2+}}$ and $\mathrm{RhoGTP}$}}
\centering
\scriptsize
\begin{tabular}{|l|l|l|p{2in}|}
\hline
        &   \multicolumn{2}{|c|}{$\mathbf{\mathrm{d_{1,2}}}$} &\\
\hline
{\bf k}       & $\mathbf{\mathrm{Ca^{2+}}}$ & $\mathbf{\mathrm{RhoGTP}}$& {\bf FUNCTIONAL ROLE}  \\
\hline
1	&0.091	 	&0.082	 	&	 PAR1 activation by SFLLRN\\
\hline
2	&0.030	 	&0.027	 	&	\\
\hline
\bf 3	&\bf 0.102	 	&\bf 0.100	 	&	\\
\hline
\bf 7	&\bf 0.218	 	&\bf 0.280	 	&	PAR1 deactivation\\
\hline
\bf 8	&\bf 0.160	 	&\bf 0.120	 	&	 PAR1 binding Gq\\
\hline
9	&0.063	 	&0.035	 	&	\\
\hline
10	&0.073	 	&0.014	 	&	Gq releasing GDP\\
\hline
11	&0.026	 	&0.012	 	&	\\
\hline
12	&0.028	 	&0.026	 	&	 Gq binding GTP\\
\hline
13	&0.054	 	&0.016	 	&	\\
\hline
\bf 14	&\bf 0.138	 	&0.055	 	&	Gq releasing $\beta\gamma$\\
\hline
15	&0.024	 	&0.017	 	&	 \\
\hline
16	&0.027	 	&0.021	 	&	Gq deactivation\\
\hline
17	&0.020	 	&0.036	 	&	${\rm PLC\beta}$ binding GqGTP\\
\hline
18	&0.038	 	&0.047	 	&	 \\
\hline
\bf 19	&\bf 0.404	 	&0.030	 	&	${\rm PLC\beta}$ hydrolizing GqGTP\\
\hline
20	&0.029	 	&0.016	 	&	${\rm PLC\beta}$ releasing GqGDP\\
\hline
21	&0.032	 	&0.035	 	&	\\
\hline
\bf 22	&\bf 0.438	 	&0.019	 	&	${\rm PLC\beta}$ binding PIP2\\
\hline
23	&0.038	 	&0.019	 	&	\\
\hline
24	&0.014	 	&0.017	 	&	${\rm PLC\beta}$ hydrolizing PIP2\\
\hline
\bf 25	&\bf 0.386	 	&0.022	 	&	 Consumption of IP3\\
\hline
\bf 26	&0.041	 	&\bf 0.375	 	&	PAR1 binding G13\\
\hline
\bf 27	&0.017	 	&\bf 0.112	 	&	\\
\hline
28	&0.028	 	&0.059	 	&	G13 releasing GDP\\
\hline
29	&0.031	 	&0.018	 	&	\\
\hline
30	&0.022	 	&0.025	 	&	G13 binding GTP\\
\hline
31	&0.019	 	&0.021	 	&	\\
\hline
32	&0.021	 	&0.019	 	&	G13 releasing $\beta\gamma$\\
\hline
33	&0.018	 	&0.038	 	&	\\
\hline
34	&0.038	 	&0.077	 	&	G13 deactivation\\
\hline
\bf 35	&0.023	 	&\bf 0.160	 	&	 GEF binding G13GTP\\
\hline
\bf 36	&0.027	 	&\bf 0.206	 	&	\\
\hline
\bf 37	&0.022	 	&\bf 0.130	 	&	GEF hydrolizing G13GTP\\
\hline
38	&0.036	 	&0.038	 	&	GEF releasing G13GDP\\
\hline
39	&0.024	 	&0.022	 	&	\\
\hline
\bf 40	&0.037	 	&\bf 0.370	 	&	 GEFG13GTP binding RhoGDP\\
\hline
41	&0.021	 	&0.096	 	&	\\
\hline
42	&0.024	 	&0.035	 	&	Rho activation\\
\hline
\bf 43	&0.020	 	&\bf 0.344	 	&	  Rho deactivation\\
\hline
\end{tabular}
\normalsize
\begin{flushleft}
The sensitivity coefficients measured as  Kolmogorov-Smirnov distances $d_{1,2}$ estimated in the ranges $[1/10,10]~\overline k_j$
for the peak values of $y_i=\mathrm{Ca^{2+}}$ (column two) and $y_i=\mathrm{RhoGTP}$ (column three).
Sensitivity coefficients larger than $10\%$ are in bold font.
The objective function used in the simulations is defined in~(\ref{eq:obj-pk}) 
and the peak values are defined in~(\ref{eq:value-pk}).
\end{flushleft}
\label{tab:tableV}
\end{table}

%% file: tableVI.tex
\begin{table}[!ht]
\caption{
\bf{The sensitivity coefficients calculated by Monte Carlo filtering for the peak times 
$\tpk^{peak}$ of $\mathrm{Ca^{2+}}$ and $\mathrm{RhoGTP}$}} 
\scriptsize
\centering
\begin{tabular}{|l|l|l|p{2in}|}
\hline
        &   \multicolumn{2}{|c|}{$\mathbf{\mathrm{d_{1,2}}}$} &\\
\hline
{\bf k}       & $\mathbf{\mathrm{Ca^{2+}}}$ & $\mathbf{\mathrm{RhoGTP}}$& {\bf FUNCTIONAL ROLE}  \\
\hline
1	&0.075	 	&0.043	 	&	 PAR1 activation by SFLLRN\\
\hline
2	&0.032	 	&0.028	 	&	\\
\hline
3	&0.031	 	&0.034	 	&	\\
\hline
\bf 7	&\bf 0.201	 	&\bf 0.530	 	&	PAR1 deactivation \\
\hline
\bf 8	&\bf 0.134	 	&\bf 0.441	 	&	 PAR1 binding Gq\\
\hline
\bf 9	&0.063	 	&\bf 0.161	 	&	\\
\hline
10	&0.081	 	&0.076	 	&	Gq releasing GDP\\
\hline
11	&0.071	 	&0.032	 	&	\\
\hline
12	&0.074	 	&0.032	 	&	 Gq binding GTP\\
\hline
13	&0.033	 	&0.045	 	&	\\
\hline
\bf 14	&\bf 0.199	 	&\bf 0.338	 	&	Gq releasing $\beta\gamma$\\
\hline
15	&0.067	 	&0.045	 	&	 \\
\hline
16	&0.076	 	&0.023	 	&	Gq deactivation\\
\hline
17	&0.052	 	&0.042	 	&	${\rm PLC\beta}$ binding GqGTP\\
\hline
18	&0.033	 	&0.032	 	&	 \\
\hline
\bf 19	&\bf 0.148	 	&0.039	 	&	${\rm PLC\beta}$ hydrolizing GqGTP\\
\hline
20	&0.025	 	&0.040	 	&	${\rm PLC\beta}$ releasing GqGDP\\
\hline
21	&0.035	 	&0.046	 	&	\\
\hline
22	&\bf 0.324	 	&0.039	 	&	${\rm PLC\beta}$ binding PIP2\\
\hline
23	&0.051	 	&0.040	 	&	\\
\hline
24	&0.066	 	&0.040	 	&	${\rm PLC\beta}$ hydrolizing PIP2\\
\hline
\bf 25	&\bf 0.318	 	&0.022	 	&	 Consumption of IP3\\
\hline
26	&0.038	 	&0.083	 	&	PAR1 binding G13\\
\hline
27	&0.023	 	&0.039	 	&	\\
\hline
28	&0.032	 	&0.059	 	&	G13 releasing GDP\\
\hline
29	&0.040  	&0.030	 	&	\\
\hline
30	&0.027	 	&0.046	 	&	G13 binding GTP\\
\hline
31	&0.052	 	&0.016	 	&	\\
\hline
32	&0.056	 	&0.027	 	&	G13 releasing $\beta\gamma$\\
\hline
33	&0.047	 	&0.065	 	&	\\
\hline
\bf 34	&0.067	 	&\bf 0.314	 	&	G13 deactivation\\
\hline
35	&0.044	 	&0.063	 	&	 GEF binding G13GTP\\
\hline
36	&0.064	 	&0.083	 	&	\\
\hline
\bf 37	&0.034	 	&\bf 0.448	 	&	GEF hydrolizing G13GTP\\
\hline
38	&0.029	 	&0.044	 	&	GEF releasing G13GDP\\
\hline
39	&0.030	 	&0.042	 	&	\\
\hline
40	&0.067	 	&0.045	 	&	 GEFG13GTP binding RhoGDP\\
\hline
41	&0.046	 	&0.051	 	&	\\
\hline
42	&0.060	 	&0.034	 	&	Rho activation\\
\hline
43	&0.029	 	&0.059	 	&	  Rho deactivation\\
\hline
\end{tabular}
\normalsize
\begin{flushleft}
The sensitivity coefficients measured as Kolmogorov-Smirnov distances $d_{1,2}$ 
estimated in the ranges $[1/10,10]~\overline k_j$ for the peak times $\tpk^{peak}$ 
of $y_i=\mathrm{Ca^{2+}}$ (column two) and $y_i=\mathrm{RhoGTP}$ (column three). 
Sensitivity coefficients larger than $10\%$ are in bold font.
The objective function used in the simulations is defined in~(\ref{eq:obj-pk}) 
and the peak times $\tpk^{peak}$ are defined in~(\ref{eq:tpk-peak}).
\end{flushleft}
\label{tab:tableVI}
\end{table}

%% file: tableVII.tex
\begin{sidewaystable}
\caption{
\bf{List of the most important parameters as deemed by the six different sensitivity tests adopted}}
{\scriptsize 
\begin{tabular}{|c|c|c|c|c|c|c|c|c|c|c|c|c|p{1.8in}|}
\hline
              &  \multicolumn{12}{|c|}{\textbf{Sensitivity Anlysis Method}} &\\
\hline
              &  \multicolumn{2}{|c|}{Taylor $t_i$} & \multicolumn{2}{|c|}{Filtering $t_i$} & \multicolumn{2}{|c|}{Taylor $T$} & \multicolumn{2}{|c|}{Filtering $T$} & \multicolumn{2}{|c|}{Filtering $y_{max}$}  & \multicolumn{2}{|c|}{Filtering $t_{max}$}& \\
\hline
$\mathrm{k}$  &  $\mathbf{\mathrm{Ca^{2+}}}$ & $\mathbf{\mathrm{RhoGTP}}$ & $\mathbf{\mathrm{Ca^{2+}}}$ & $\mathbf{\mathrm{RhoGTP}}$ & $\mathbf{\mathrm{Ca^{2+}}}$ & $\mathbf{\mathrm{RhoGTP}}$ & $\mathbf{\mathrm{Ca^{2+}}}$ & $\mathbf{\mathrm{RhoGTP}}$ & $\mathbf{\mathrm{Ca^{2+}}}$ & $\mathbf{\mathrm{RhoGTP}}$ & $\mathbf{\mathrm{Ca^{2+}}}$ & $\mathbf{\mathrm{RhoGTP}}$ & \textbf{FUNCTIONAL ROLE}\\ 
\hline
3 &  & &$\times$ & & &  & & &$\times$  &$\times$ &  & &Consumption PAR1-AP\\
\hline
7 &  &$\times$ &$\times$ &$\times$ & & $\times$ &$\times$ &$\times$ & $\times$ &$\times$ & $\times$ &$\times$ & PAR1 deactivation\\  
\hline
8 &  & & &$\times$ & &  &$\times$ & & $\times$ &$\times$ & $\times$ &$\times$ &PAR1 binding Gq\\
\hline
9 &  & & & & &  & & &  & &  & $\times$&\\
\hline
14 &  & &$\times$ & & &  & & &$\times$  & &$\times$  &$\times$ &Gq releasing $\beta\gamma$\\
\hline
19 &  & &$\times$ & & &  &$\times$ & &$\times$  & &$\times$  & &PLC$\beta$ hydrolyzing GqGTP\\
\hline
22 &  & &$\times$ & & &  &$\times$ & &$\times$  & & $\times$ & &PLC$\beta$ binding PIP2\\
\hline
25 &  & &$\times$ & & &  &$\times$ & &$\times$  & &$\times$  & &Consumption of IP3\\
\hline
26 &  &$\times$ & &$\times$ & &$\times$  & &$\times$ &  &$\times$ &  & &PAR1 binding G13\\
\hline
27 &  & & &$\times$ & &  & & &  &$\times$ &  & &\\
\hline
34 &  & & & & &$\times$  & & &  & &  &$\times$ &G13 deactivation\\
\hline
35 &  &$\times$ & &$\times$ & &  & & &  &$\times$ &  & &GEF binding G13GTP\\
\hline
36 &  &$\times$ & &$\times$ & &  & &$\times$ &  &$\times$ &  & &\\
\hline
37 &  & & &$\times$ & &$\times$  & &$\times$ &  &$\times$ &  &$\times$ &GEF hydrolizing G13GTP\\
\hline
40 &  &$\times$ & &$\times$ & &$\times$  & &$\times$ &  &$\times$ &  & &GEFG13* binding Rho\\
\hline
41 &  & & &$\times$ & &  & & &  & &  & &\\
\hline
43 &  &$\times$ & &$\times$ & &$\times$  & &$\times$ &  & $\times$&  & &Rho deactivation\\
\hline
\end{tabular}
}
\begin{flushleft}
Comparison of the sensitivity analysis results for the most important parameters reported in bold font
in Tables~\ref{tab:tableI}--\ref{tab:tableVI}. The symbols $\times$ mark the importance of a given parameter $k_j$ 
according to the different sensitivity test. Parameters $k_7$ and $k_8$ are important in the majority of the sensitivity tests
performed.  
\end{flushleft}
\label{tab:tableVII}
\end{sidewaystable}